# Thermal expansion anisotropy of the $Fe_{23}Mo_{16}$ and $Fe_7Mo_6$ μ-phases predicted from first-principles calculations

**Dmitry Vasilyev**

Baikov Institute of Metallurgy and Materials Science of RAS, 119334, Moscow, Leninsky Prospekt 49, Russia

dvasilyev@imet.ac.ru ; vasilyev-d@yandex.ru

**Abstract**. The intermetallic μ- phase, which precipitates in steels and superalloys, can noticeably soften the mechanical properties of their matrix. Despite the importance of developing superalloys and steels, the thermodynamic properties and directions of thermal expansion of the μ-phase are still poorly studied. In this work, the thermal expansion paths, elastic, thermal and thermodynamic properties of the $Fe_{23}Mo_{16}$ and $Fe_7Mo_6$ μ-phases have been studied using first-principles based quasi-harmonic Debye - Grüneisen approach. A method allowing avoids differentiation in many variables is used. The free energies consisting of the electronic, vibrational and magnetic energy contributions, calculated along different paths of thermal expansions were compared between themselves. A path with the least free energy was chosen as the trajectory of thermal expansion. Negative thermal expansion of the $Fe_7Mo_6$ compound was predicted, while the $Fe_{23}Mo_{16}$ has a conventional thermal expansion. The thermal expansions of both these compounds are not isotropic. The elastic constants, modulus, heat capacities, Curie and Debye temperatures were predicted. The obtained results satisfactorily agree with the available experimental data. Physical factors affecting the stability of $Fe_{23}Mo_{16}$ and $Fe_7Mo_6$ have been studied. The paper presents an essential feature of thermal expansions of the μ-phase of the Fe-Mo system, which can provide an insight into future developments.

## 1. Introduction

The μ-phase belongs to the class of topologically close-packed (TCP) phases and is an undesirable precipitation in high-doped superalloys and various types of steels, such as nickel-based or ferritic-martensitic steels. It is generally believed that the precipitation of the μ-phase degrades the mechanical properties of these materials, in particular, affects the impact toughness, creep, corrosion resistances, yield strength, and ductility [1-3].

Superalloys are used in production of turbine blades for aircraft engines, while ferritic steels applications are found in developing a fuel element cladding of the new generation reactors and considering as a promising material for the first wall of future fusion reactors. It's due to excellent high-temperature mechanical properties, creep and corrosion resistance, and low swelling under irradiation [4, 5].

In the Fe-Mo phase diagram according to Massalski et al. [6], the μ-phase has a homogeneity range from approximately 38 to 44 at. % Mo. Thus, the $Fe_7Mo_6$ compound with a stoichiometric composition of 46.2 at.% Mo is considered as a metastable phase.

Nevertheless, the knowledge of thermodynamic properties of $Fe_7Mo_6$ is important, since $Fe_7Mo_6$ is one of model phases used to study the precipitation of TCP phases in ferritic steels and superalloys as a result of long-term operation of products at high temperatures and irradiation. Therefore, $Fe_7Mo_6$ attracts a lot of attention and was studied in many works. But, experimental studies of metastable compounds are difficult due to diffusion during long-term exposure of materials at heat treatment temperatures, and a possible deviation in the concentration of elements from the initial stoichiometry. Therefore, it is a reasonable idea to carry out calculations for a known stable compound close in composition to the model μ-phase and compare the obtained results between themselves and with the experimental data. The $Fe_{23}Mo_{16}$ μ-phase has a stoichiometry composition of 41 at.% Mo, and is stable, therefore this compound was chosen for comparative analysis.

In recent decades, a large number of works devoted to the study of the μ-phase have been carried out. Lejeaghere et al. [7] obtained optimized atoms positions of $Fe_7Mo_6$ and total magnetic moments with the help of the electronic density functional theory (DFT). Cieslak et al. [8] studied structural and electronic properties of the μ-phase of Fe-Mo compounds experimentally by X-ray diffraction and Mossbauer spectroscopy. The structure properties of μ-phases were investigated by Sluiter et al. [9]. The study of metastable μ-phases in the Fe – Mo system was carried out using DFT in [10]. Rajkumar et al. [11] predicted the lattice parameters of $Fe_7Mo_6$ compound by DFT calculations. Schroders et al. [12] studied the plastic deformation and the structure of mobile defects in the μ-phase by transmission electron microscopy, the samples with the stoichiometry of $Fe_7Mo_6$ were annealed at T = 800 C for 1000 hours. Eumann et al. [13] reported the phase equilibria of μ-phases in binary and ternary alloys at T = 800 C after annealing for 1500 - 2000 hours, from metallography, X-ray diffraction and electron probe microanalysis. Experimental studies were carried out to determine the structure and lattice parameters of μ-phases [14-17]. Precipitations of the μ-phase in nickel-based alloys were studied by scanning, transmission electron microscopy, and X-ray diffraction by Zhao et al. [18]. A study of the μ-phases precipitated in superalloys at T = 800 C, where the thermal treatment reached 10000 hours, was performed by Qin et al. [19]. Studies of the distribution of atoms over different sublattices, the structure and lattice parameters of μ-phases were carried out in [20, 21]. The phase diagrams of Co–Mo and Fe–Mo systems by means of a combination of ab initio electronic structure calculations and the CALPHAD approach were calculated by Houserová et al. [22]. The formation of standard Gibbs energies for compounds of the Fe-Mo-O system was experimentally obtained by Koyama et al. [23].

Thus, most of these works are focused on the structural properties of μ-phases. But the thermodynamic and elastic properties of $Fe_{23}Mo_{16}$ and $Fe_7Mo_6$ are not compared. The causes of stabilizing these compounds have not been studied so far. But taking into account the magnetic entropy can significantly increase the stability of alloys, as reported in [24]. It was shown in [25, 26] that accounting the magnetic and electronic subsystems is a necessary condition for correct thermodynamic description of compounds containing elements with magnetic properties. In addition, the direction of thermal expansion pathways of μ-phase is not studied. Moreover, the tensile behavior and impact toughness of superalloys after precipitation of the μ-phase still raises questions, as noted, for example, in [18, 19, 27].

Since these data are essential for obtaining Gibbs energies, calculating the thermodynamics of μ-phases and computing boundaries of the Fe-Mo phase diagram, this work was carried out.

First-principles calculations based on density functional theory (DFT) combined with the quasi-harmonic Debye–Grüneisen (QDG) approximation [28] were used. A method of searching a thermal expansion path (STEP) [24, 25] was employed in this work. This method allows to simplify calculations by reducing the problem from a two-dimensional task to a one-dimensional case.

Calculations in this work are limited by the stoichiometry of the $Fe_{23}Mo_{16}$ and $Fe_7Mo_6$ compounds; possible deviations from stoichiometry were not taken into account.

The anisotropy of thermal expansion paths of $Fe_{23}Mo_{16}$ and $Fe_7Mo_6$ compounds was studied by comparing free energies calculated along different directions of thermal expansion. It was predicted that the metastable $Fe_7Mo_6$ compound has a negative thermal expansion, and the stable $Fe_{23}Mo_{16}$ has a conventional one, while having a negative thermal expansion coefficient (TEC) in parameter *c*. The nature of thermal expansions of these both compounds is not isotropic.

Electronic, vibrational and magnetic energy subsystems were taking into account for description of free energy of compounds. Elastic, thermal and thermodynamic properties, sound wave velocities, Curie and Debye temperatures of $Fe_{23}Mo_{16}$ and $Fe_7Mo_6$ were predicted.

The obtained results can bring us closer to understanding the mechanism of embrittlement of the superalloy matrix during precipitation of the µ-phase. And can be useful for constructing Gibbs potentials of µ-phase and further refining of the Fe-Mo phase diagram boundaries. What is a necessary condition for the further design of superalloys and ferritic steels.

## 2. Methodology

### 2.1. Ab initio calculations

The total static energies of $Fe_{23}Mo_{16}$ and $Fe_7Mo_6$ were calculated with the spin-polarized calculations using the FP-(L)APW + lo method implemented in WIEN2k code [29]. It was used for defining the structural, elastic and electronic properties. The values of 2.05 and 2.15 (a.u.) for the RMT radii of Fe and Mo atoms were used in these calculations. R·$K_{max}$ was set as equal to 8. GGA-PBE [30] was used for the exchange-correlation potential. The mesh of the Brillouin zone [31] depends on the size of the unit cells and the number of symmetry operations which can be imposed on the lattice. In this work, the mesh of 15 x 15 x 3 *k*-point was used. The total energy of the SCF calculations was set for an energy convergence of $10^{-6}$ eV/atom.

### 2.2. Free energy models

Helmholtz free energy *F* as a function of volume (*V*) and temperature (*T*) can be computed within the adiabatic approximation as follows [32]

$$F(V,T) = E_{stat}(V) + F_{el}(V,T) + F_{vib}(V,T) + F_{mag}(V,T) - TS_{conf} \qquad (1)$$

where $E_{stat}(V)$ is the total static energy calculated by DFT at the ground state. The rest parts of the equation are the electronic, vibrational, magnetic energies contributions, and $S_{conf}$ is the ideal configurational entropy.

The thermal electron excitation energy $F_{el}(V,T)$ to the free energy was formulated as in [33]

$$F_{el}(V,T) = E_{el}(V,T) - TS_{el}(V,T) \qquad (2)$$

Where the electronic entropy $S_{el}(V,T)$ is formulated by

$$S_{el}(V,T) = -k_B \int_{-\infty}^{\infty} n(\varepsilon,V)\big(f(\varepsilon,T)\ln f(\varepsilon,T) + \big(1 - f(\varepsilon,T)\big)\ln(1 - f(\varepsilon,T))\big)d\varepsilon \qquad (3)$$

With the energy of electrons $E_{el}(V,T)$ takes the form

$$E_{el}(V,T) = \int_{-\infty}^{\infty} n(\varepsilon,V)f(\varepsilon,T)\varepsilon d\varepsilon - \int_{-\infty}^{\varepsilon_F} n(\varepsilon,V)\varepsilon d\varepsilon \qquad (4)$$

where $n(\varepsilon,V)$ is the total electronic density of states DOS, $f(\varepsilon,T)$ is the Fermi-Dirac distribution, $\varepsilon_F$ is Fermi energy, $k_B$ is the Boltzmann constant.

The quasi-harmonic Debye- Grüneisen approximation was used to describe the free energy of lattice ion vibrations as given in [28]

$$F_{vib}(V,T) = E_{vib}(V,T) - TS_{vib}(V,T), \qquad (5)$$

where the energy of ions vibrations $E_{vib}(V,T)$ were expressed as

$$E_{vib}(T,V) = \frac{9}{8}N_A k_B \theta_D + 3N_A k_B T D\left(\frac{\theta_D}{T}\right) \qquad (6)$$

while, the vibrational entropy $S_{vib}(T,V)$ takes the form

$$S_{vib}(T,V) = 3N_A k_B \left[\frac{4}{3}D\left(\frac{\theta_D}{T}\right) - ln\left(1 - exp\left(-\frac{\theta_D}{T}\right)\right)\right] \qquad (7)$$

where, $D(\theta_D/T)$ is Debye function; $\gamma$ – the Grüneisen parameter was formulated as follows

$$\gamma = -1 - \frac{V}{2}\frac{\partial^2 P/\partial V^2}{\partial P/\partial V} \qquad (8)$$

Debye temperature $\theta_D(V)$ was expressed as

$$\theta_D(V) = \theta_{Do}\left(\frac{V_0}{V}\right)^{\gamma} \qquad (9)$$

where, the Debye temperature $\theta_{D0}$ was calculated at the ground state at equilibrium volume $V_0$; the $V_0$ and $\theta_{D0}$ are presented in Table 1 and Table 4.

Since Fe atoms are elements with magnetic properties, µ-phase should have a ferro/paramagnetic transition. In order to take into account this phase transition, the Hillert – Jarl (HJ) model was used according to [34]

$$F'_{mag}(T) = RT Ln(\beta + 1)f(\tau) \qquad (10)$$

the function $f(\tau)$ is defined in HJ model, where $\tau$ equals to $T/T_c$ ratio, $T_C$ - Cure temperature, $\beta$ - the average magnetic moment of atoms.

Curie temperature can be estimated using the mean-field approximation as follows

$$T_C = \frac{2}{3k_B}(E_{stat}^{PM}(V_0) - E_{stat}^{FM}(V_0)) \qquad (11)$$

where $E^{PM}_{stat}(V_0)$ and $E^{FM}_{stat}(V_0)$ are the total static energies calculated by DFT at the ground state for paramagnetic (PM) and ferromagnetic (FM) states of the compound at the corresponding equilibrium volumes $V_0$.

In order to compensate the constant contribution of magnetic internal energy which persistently exist in the total static energy calculations the magnetic free energy $F_{mag}(V,T)$ was formulated in accordance with the following assumption

$$F_{mag}(V,T) = [F'_{mag}(T) - F'_{mag}(0K)] - TS_{mag}(V) \qquad (12)$$

Magnetic $S_{mag}(V)$ and configurational $S_{conf}$ entropies were formulated in accordance with expressions

$$S_{mag}(V) = k_B \sum_{i=1}^{n} c_i ln\left(|\mu_i(V)| + 1\right) \qquad (13)$$

$$S_{conf} = k_B \sum_{i=1}^{n} c_i ln\, c_i \qquad (14)$$

where $\mu_i$ and $c_i$ – the local magnetic moment and concentration of atom $i$, respectively.

## 3. Results and discussion

### 3.1. Ground state properties

#### 3.1.1. Phase stability of compounds

The crystal structure of $Fe_7Mo_6$ µ-phase is typified by $Fe_7W_6$ compound, which crystallizes in the space group *R3m* (No. 166), is shown in Figure 1 (a) with 39 atoms per unit cell containing 21 Fe atoms and 18 Mo atoms.

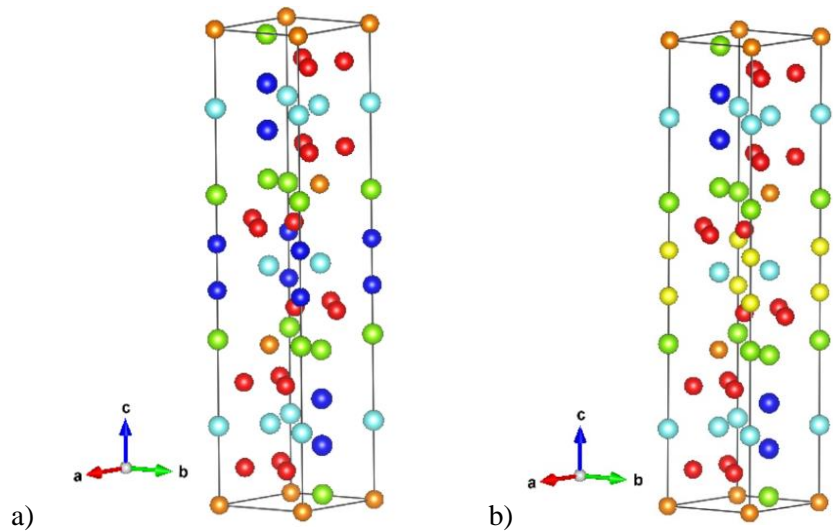

**Figure 1**. The crystal lattice of the µ-phase. a) In the $Fe_7Mo_6$ unit cell, Fe atoms occupy the *3a* and *18h* Wyckoff sites are shown in orange and red respectively, and Mo atoms occupy the *6c*, *6c'* and *6c''* sites are shown in light blue, bright green and blue respectively; b) while, in the $Fe_{23}Mo_{16}$ unit cell the *6c''* sites are occupied by two Fe atoms shown in yellow and four Mo atoms shown in blue.

The $Fe_{23}Mo_{16}$ compound has 21 iron atoms occupying the *3a* and *18h* Wyckoff sites as shown in Figure 1 (b) in orange and red respectively; additional two iron atoms can occupy along with the sixteen molybdenum atoms any of the *6c*, *6c'* and *6c''* sites which shown in Figure 1 (b), where the *6c* and *6c'* sites are shown in light blue and bright green, while the *6c''* sites are shown in blue and yellow respectively.

After optimisation procedures the optimal atoms distribution on the sub-lattices of the $Fe_{23}Mo_{16}$ was obtained. It turned out that the most energetically favorable position is when the *6c''* sub-lattice is occupied by two Fe atoms and four Mo atoms, as shown in yellow and blue in Figure 1 (b), respectively. Thus, this configuration was used in these calculations.

The lattice parameters and atoms positions of $Fe_{23}Mo_{16}$ and $Fe_7Mo_6$ compounds were optimized with the spin-polarized calculations by the geometry optimizations and the full structural relaxation, which were carried out from ideal positions.

The optimized lattice parameters are listed in Table 1, Figure 3 (a) and Figure 3 (b) together with the available experimental data [12, 13, 16] and other theoretical results [11] for comparison. As follows from Table 1 and Figure 3 (b), the calculated parameters of the equilibrium lattice of $Fe_7Mo_6$ are in a satisfactory agreement, within 1%, with the theoretical data [11] calculated at T = 0 K. The enthalpy formation $\Delta H$ of $Fe_{23}Mo_{16}$ and $Fe_7Mo_6$ was formulated as

$$\Delta H^{AB} = E_{stat}^{AB} - \left(xE_{solid}^{A} + (1-x)E_{solid}^{B}\right) \qquad (15)$$

where $E^{AB}_{stat}$ is the total static energy of the unit sell of the Fe$_{23}$Mo$_{16}$ and Fe$_7$Mo$_6$, $E^{A}_{solid}$ and $E^{B}_{solid}$ are the energies of Fe and Mo atoms in the solid state, i.e. in the body centred cubic (bcc) lattice; $x$ is the concentration of Mo atoms ($x = 0.41$, or 0.462).

The calculated negative values ΔH of Fe$_{23}$Mo$_{16}$ and Fe$_7$Mo$_6$ obtained in this work, which are listed in Table 1 with the other available theoretical values for comparison, mean that these compounds are stable at the ground state. The ΔH of Fe$_7$Mo$_6$ calculated in [35] has the positive value at T = 0 K with respect to Fe and Mo, perhaps such a result can be obtained if Fe$_7$Mo$_6$ is considered to be in the nonmagnetic state.

**Table 1**
The lattice parameters (in Å), equilibrium volume V (in Å$^3$), formation enthalpies ΔH (kJ/mol) of Fe$_{23}$Mo$_{16}$ and Fe$_7$Mo$_6$ calculated in this research and obtained in other theoretical and experimental works.

| Compound | Method | a | c | c/a | V | ΔH |
|---|---|---|---|---|---|---|
| Fe$_7$Mo$_6$ | This work | 4.765 | 25.768 | 5.408 | 506.661 | -2.525 |
| | This work [1)] | 4.677 | 25.827 | 5.522 | 489.305 | |
| | calc. [11] | 4.766 | 25.807 | 5.415 | 507.664 | -0.352 |
| | exp. [16] | 4.751 | 25.680 | 5.405 | 501.991 | |
| | exp. [13] [2)] | 4.73 | 25.85 | 5.465 | 500.857 | |
| | exp. [13] [3)] | 4.757 | 25.710 | 5.405 | 503.847 | |
| | exp. [17] | 4.75 | 25.72 | 5.415 | 502.561 | |
| | exp. [12] [4)] | 4.757 | 25.753 | 5.414 | 504.690 | |
| | ref. [11] | | | | | -1.418 |
| | calc. [35] | | | | | +2.7 |
| Fe$_{23}$Mo$_{16}$ | This work | 4.737 | 25.743 | 5.434 | 500.259 | -1.03 |
| | This work [1)] | 4.780 | 25.652 | 5.366 | 507.702 | |

[1)] calculated in this work at T = 1073 K.
Experimental data obtained by annealing treatments at T = 1073 K after:
[2)] 1500 hours.
[3)] 2000 hours.
[4)] 1000 hours by employing the Rietveld method; samples of the stoichiometric composition of Fe$_7$Mo$_6$ were prepared by arc-melting.

*3.1.2. Elastic and thermal properties of Fe$_{23}$Mo$_{16}$ and Fe$_7$Mo$_6$ compounds*

The strain tensor coefficients $C_{ij}$ for Fe$_{23}$Mo$_{16}$ and Fe$_7$Mo$_6$ have been calculated with the help of distortion matrices $\boldsymbol{D_i}$ given in Table A1 (in Appendix), which were successively imposed on the lattices. In effect, $C_{ij}$ were calculated from the total energy changes $\Delta E$ between the initial $\boldsymbol{R}$ (A1) and deformed $\boldsymbol{R'}$ lattices, according to the matrix equation: $\boldsymbol{R}\cdot\boldsymbol{D_i} = \boldsymbol{R'}$. The energy differences $\Delta E(\delta) = E_{stat}(V, \delta_i) - E_{stat}(V_0,0)$ for the various distortions $\boldsymbol{D_i}$ applied to lattices are shown in Figure 2 (a) and Figure 2 (b).

Since it is known that the hexagonal crystal lattice of the μ-phase has five independent coefficients of the strain tensor [36], therefore, five independent and one dependent ($C_{66}=(C_{11}-C_{12})/2$) elastic constants $C_{ij}$ were computed by solving the corresponding set of equations $\Delta E(\delta)$ listed in Table A1. Table 2 presents the calculated elastic constants $C_{ij}$ obtained in this research and in the theoretical work [35] for comparison.

The mechanical stability of Fe$_{23}$Mo$_{16}$ and Fe$_7$Mo$_6$ can be confirmed with the help of $C_{ij}$ constants using the criterion [36], which imposes the following restrictions on the strain tensor constants of hexagonal crystals at zero pressure

$$C_{11} > 0; (C_{11} \cdot C_{33} - 2C_{13}^2 + C_{12} \cdot C_{33}) > 0; C_{11} - |C_{12}| > 0; C_{44} > 0 \qquad (16)$$

**Table 2**
The elastic constants $C_{ij}$ (in GPa) of $Fe_{23}Mo_{16}$ and $Fe_7Mo_6$ calculated in this research are shown in comparison with $C_{ij}$ obtained in [35].

| Compound | Method | $C_{11}$ | $C_{12}$ | $C_{13}$ | $C_{33}$ | $C_{44}$ | $C_{66}$ |
|---|---|---|---|---|---|---|---|
| $Fe_7Mo_6$ | This work | 424.57 | 191.73 | 136.27 | 393.86 | 103.55 | 116.42 |
|  | calc. [35] | 433 | 216 | 144 | 389 | 102 | 109 |
| $Fe_{23}Mo_{16}$ | This work | 417.0 | 185.6 | 132.9 | 404.7 | 98.8 | 115.7 |

Thus, as follows from Table 2, this criterion is met, hence $Fe_{23}Mo_{16}$ and $Fe_7Mo_6$ are mechanically stable at zero pressure and the ground state.

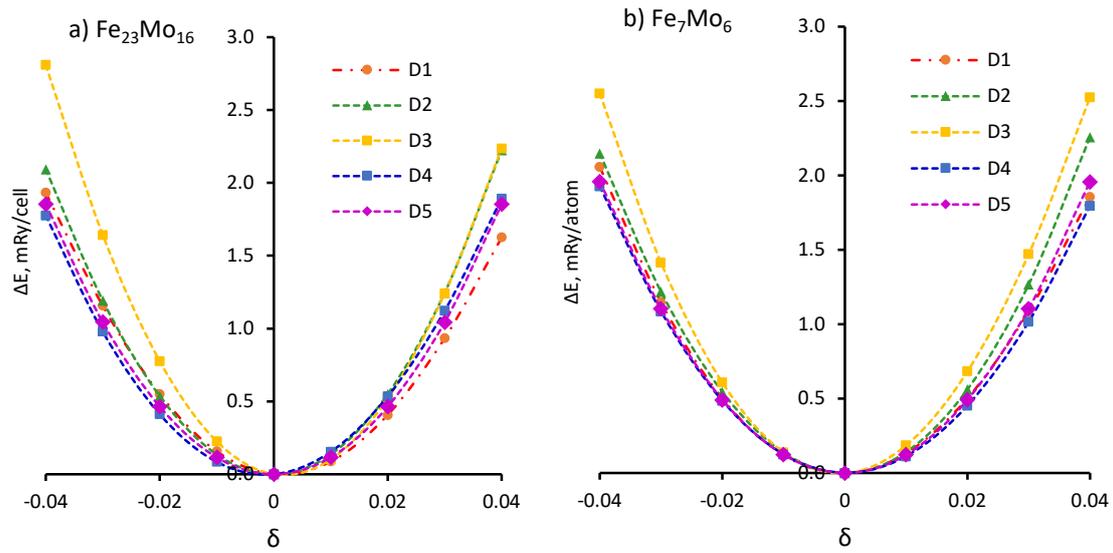

**Figure 2**. The graphs of total energy changes $\Delta E(\delta)$ as a result of applied strains after imposing the distortion matrices $D_1 – D_5$ on the lattice of a) $Fe_{23}Mo_{16}$ and b) $Fe_7Mo_6$, calculated at the ground state. The dotted lines are polynomial approximations.

From the elastic constants $C_{ij}$ obtained for a single crystal the elasticity parameters for polycrystalline aggregate were calculated by Voigt – Reuss – Hill (VRH) approximation [37]. The bulk $B$, shear $G$ moduli, Young's modulus $E$, and Poisson's ratio $v$, listed in Table 3, were obtained using the VRH approximation, by (A2 – A7). The calculated elastic compliances $S_{ij}$ are listed in Table A2.

**Table 3**
Calculated ground state elastic modulus (GPa) and Poisson's ratio $v$ for $Fe_{23}Mo_{16}$ and $Fe_7Mo_6$ in comparison with calculated values obtained in [35].

| Compound | Method | $B_V$ | $B_R$ | $B$ | $G_V$ | $G_R$ | $G$ | $E$ | $v$ | $B/G$ |
|---|---|---|---|---|---|---|---|---|---|---|
| $Fe_7Mo_6$ | This work | 241.3 | 239.4 | 240.3 | 116.7 | 99.1 | 107.8 | 281.4 | 0.30 | 2.23 |
|  | calc. [35] | 253 | 249 | 251 | 113 | 111 | 112 | - | 0.31 | 2.24 |
| $Fe_{23}Mo_{16}$ | This work | 237.9 | 236.9 | 237.4 | 115.2 | 97.2 | 106.2 | 277.2 | 0.31 | 2.24 |

To distinguish the plastic behavior of the material from the brittle one, it is customary to use values of Poisson's ratio $v = 0.26$ proposed in the study by Levandowski et al. [38] and the B/G = 1.75 suggested by Pugh [39]. If the test material has parameters greater than these indicated values, then the material is considered as ductile. Otherwise, the test material has a brittle nature. Thus, as follows from Table 3, according to the obtained values of B/G and ν, $Fe_{23}Mo_{16}$ and $Fe_7Mo_6$ are compounds with a ductile nature in the ground state.

Sound waves velocities $v_t$, $v_l$, $V_M$, Debye temperatures $\theta_D$ calculated by (A8 – A10), and anisotropic sound waves velocities obtained by (A11, A12) for $Fe_{23}Mo_{16}$ and $Fe_7Mo_6$ are given in Table 4. The Curie temperatures $T_C$ were estimated from (11), assuming that the PM state can be modelled using the nonmagnetic (NM) description, are listed in Table 4, too. This estimate of $T_C$ is relatively rough and serves to understand the behavior of the Curie temperature of compounds. From Table 4 it follows that with increasing concentration of Mo, the value of $T_C$ decreases, which corresponds to the general idea of the behavior of compounds with magnetic materials.

**Table 4**
The calculated average ($V_m$), transverse ($v_t$) and longitudinal ($v_l$) elastic wave velocities (m/s); Debye temperatures $\theta_D$ (K), Curie temperatures $T_c$ (K) and elastic wave velocities along [001] and [100] crystallographic directions (m/s) for $Fe_{23}Mo_{16}$ and $Fe_7Mo_6$.

| Compound | $v_t$ | $v_l$ | $V_m$ | $\theta_D$ | $T_C$ | [001] | | [100] | | |
|---|---|---|---|---|---|---|---|---|---|---|
| | | | | | | $V_l$ | $V_T$ | $V_l$ | $V_{T1}$ | $V_{T2}$ |
| $Fe_7Mo_6$ | 3369 | 6357 | 3765 | 477 | 193 | 6437 | 3301 | 6684 | 3500 | 3301 |
| $Fe_{23}Mo_{16}$ | 3368 | 6363 | 3765 | 474 | 272 | 6576 | 3249 | 6675 | 3516 | 3249 |

For the best knowledge, there are no experimental or theoretical studies on Debye and Curie temperatures, and sound propagation velocities for $Fe_{23}Mo_{16}$ and $Fe_7Mo_6$. Thus, additional studies are necessary to validate these results.

The fastest propagation of elastic waves, shown in Table 4, are found along the [100] direction by longitudinal waves, the similar results were obtained for Laves $Fe_2Mo$ and μ- $Co_7Mo_6$ phases reported in [24, 25]. Prediction of elastic wave velocities in different directions of the crystals may be useful for experimental studies.

### *3.2. Thermodynamic functions of $Fe_{23}Mo_{16}$ and $Fe_7Mo_6$ at finite temperatures*

#### *3.2.1. Layouts of calculations*

Comparison of the lattice parameters of $Fe_{23}Mo_{16}$ and $Fe_7Mo_6$ μ- phases between those experimentally obtained at T = 1073 K [12, 13, 16] and theoretically calculated in this work at T = 0 K shows that these compounds should have anisotropic thermal expansions. Thus, the approach of searching a thermal expansion path (STEP) [24-26] used to predict the thermal expansions of TCP Laves $Fe_2Mo$ and μ-$Co_7Mo_6$ phases was employed.

The scheme used in STEP is as follows. Let us consider the coordinate plane, where the parameters *a* and *c* of the studied crystal lattice are located along the abscissa and ordinate axes. Then we draw a number of possible directions (pathways) of thermal expansion and contraction of the compound, intersecting at the point with coordinates ($a_0$, $c_0$). This point corresponds to the equilibrium volume ($V_0$) of the compound at T = 0K, obtained after the optimization procedure. Along one of the directions, the *c/a* ratio remains constant; this is the path of isotropic expansion and contraction of the material. If experimental data are available, then it is also expedient to plot routes through the coordinates of these points. The number of such routes is arbitrary and can be increased as needed. Along each pathway we calculate the free energy using (1). A pathway with the least free energy is chosen as the path of thermal expansion.

A number of selected pathways for $Fe_{23}Mo_{16}$ and $Fe_7Mo_6$ are shown in Figure 3 (a) and Figure 3 (b). The intersection points of the $k0 \div k9$ and $p0 \div p13$ pathways correspond to the optimized lattice parameters ($a_0$, $c_0$) of $Fe_{23}Mo_{16}$ and $Fe_7Mo_6$ calculated in this work by DFT at T = 0K. The $k0$ and $p0$ directions correspond to the trajectory of isotropic expansions of $Fe_{23}Mo_6$ and $Fe_7Mo_6$, where the $c/a$ = 5.434 and 5.408 ratios remain constants. Along the $k1$ and $p10$ the $a$- parameters of both compounds remain constant. The $p1$ passes through the experimental point obtained at T = 1073K [13] and shown in Figure 3 (b) by the red triangle. As mentioned above, $Fe_7Mo_6$ is a metastable compound; therefore, the experimental data [12, 13, 16] are shown in Figure 3 (a) and Figure 3 (b) for comparison.

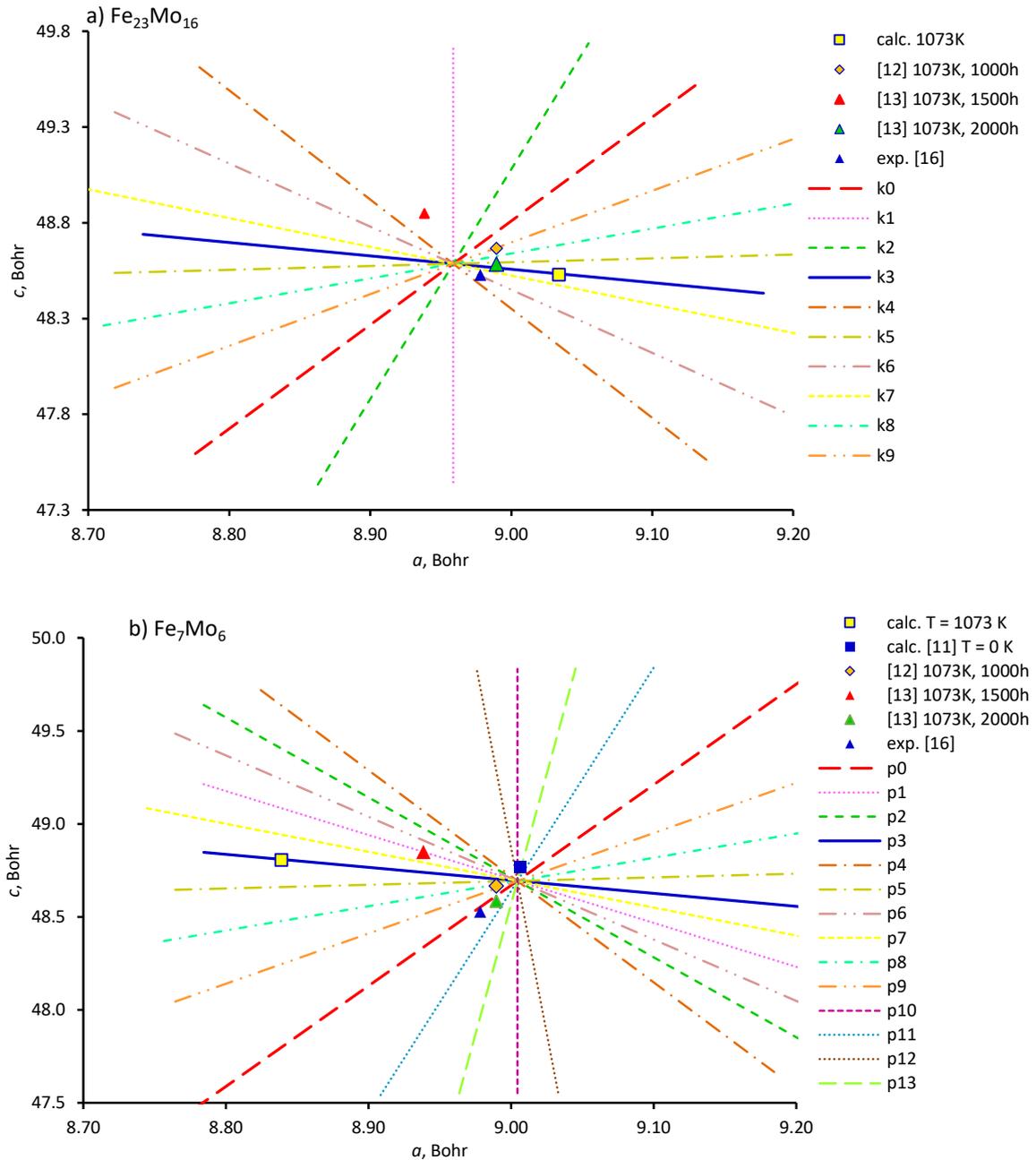

**Figure 3**. The outline of the calculations for a) $Fe_{23}Mo_{16}$ and b) $Fe_7Mo_6$ compounds. The intersection point of all routes corresponds to the optimized lattice parameters ($a_0$, $c_0$) calculated in this work by DFT at T = 0 K. The red dashed lines $k0$ and $p0$ correspond to the isotropic thermal expansion of $Fe_{23}Mo_{16}$ and $Fe_7Mo_6$, respectively; the blue solid line $k3$ is the calculated thermal expansion pathway

of $Fe_{23}Mo_{16}$, while the blue solid line *p3* is the predicted pathway of the negative thermal expansion of $Fe_7Mo_6$. The obtained lattice parameters of $Fe_{23}Mo_{16}$ and $Fe_7Mo_6$ calculated in this work at T = 1073 K are shown by yellow squares. The coordinates of lattice parameters obtained in experimental studies [12, 13, 16] at T = 1073 K and the theoretical work [11] calculated at T = 0 K, are shown for comparison.

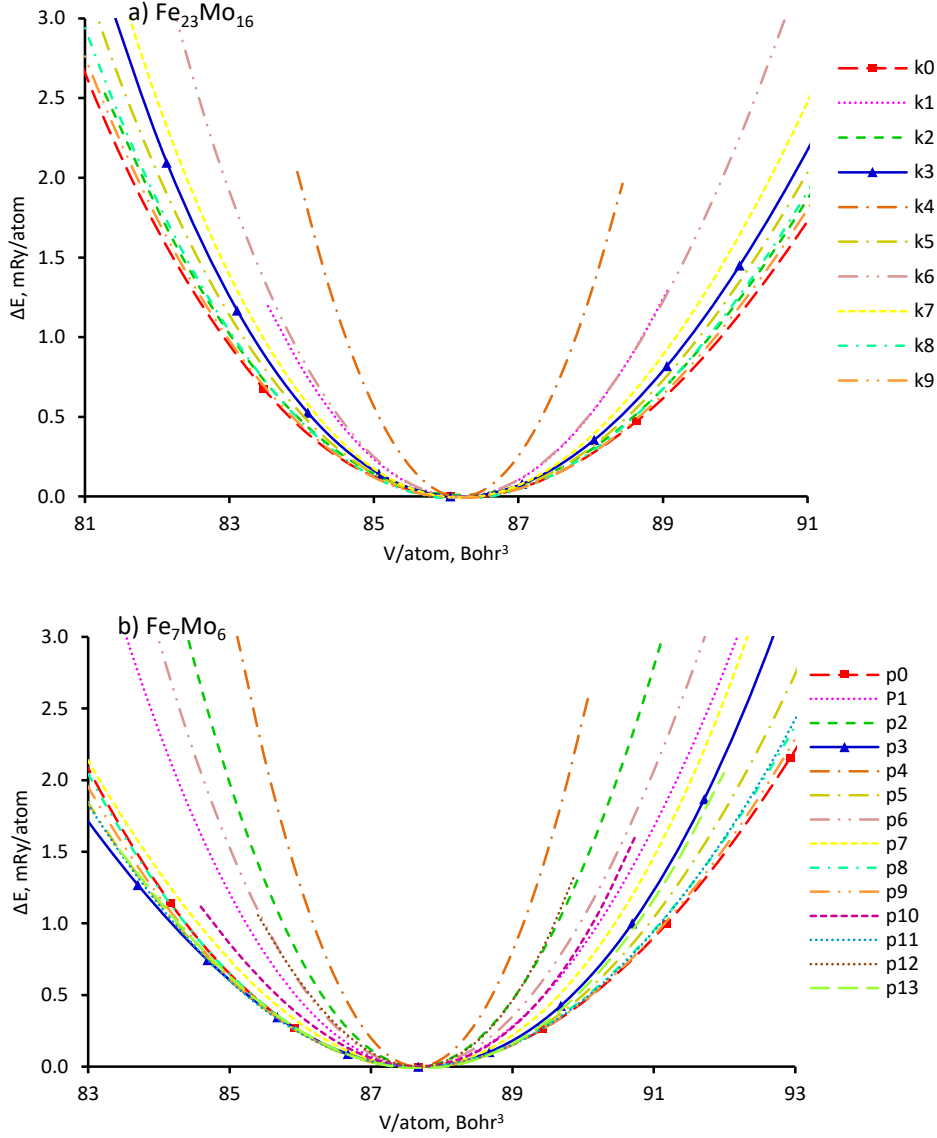

**Figure 4**. Total static energies $E_{stat}(V)$ as functions of the volume calculated by DFT for a) $Fe_{23}Mo_{16}$ along the $k0 \div k9$, and b) $Fe_7Mo_6$ along the $p0 \div p13$ pathways.

It's known that the μ-phase volume depends on *a* and *c* parameters, $V(a,c) = (\sqrt{3}/2) \cdot a^2 \cdot c$. Therefore, from a formal point of view, we must consider the total energy $E(a,c)$ as a function of two variables *a* and *c*. Thus, we have to differentiate $E(a,c)$ with respect to these two variables to get, for example, the pressure $P(a,c)$ or bulk modulus $B(a,c)$. On the other hand, the total energies $E(V)$ can be calculated along the $k0 \div k9$ and $p0 \div p13$ routes, as shown in Figure 4 (a) and Figure 4 (b), and can be considering as functions depending only on one variable, the volume. This method allows us to reduce the problem to a one-dimensional case. So we can use the Vinet equation of state (EOS) [40] and the quasi-harmonic Debye- Grüneisen (QDG) model [28] along each *ki* and *pi* routes to calculate the free energies $F(V,T)$ using (1) and simulate thermodynamic properties. Thus, if we compare the free energies $F(V,T)$ calculated along the $k0 \div k9$ and $p0 \div p13$ between themselves, we can find out the most energetically favourable thermal expansion pathway of $Fe_{23}Mo_{16}$ and $Fe_7Mo_6$.

The application of this method, in fact, gives us the formal right to use Equation (1) in this work. Because along each fixed pathways, we can use the EOS and the Debye approach [28], which are functions depending on one variable - the volume.

An accuracy of the calculations can be increased if we add more extra routes and comparing the free energies calculated along. In effect, thermal expansion of a compound may turn out to be non-linear.

### 3.2.2. Vibrational subsystem of $Fe_{23}Mo_{16}$ and $Fe_7Mo_6$

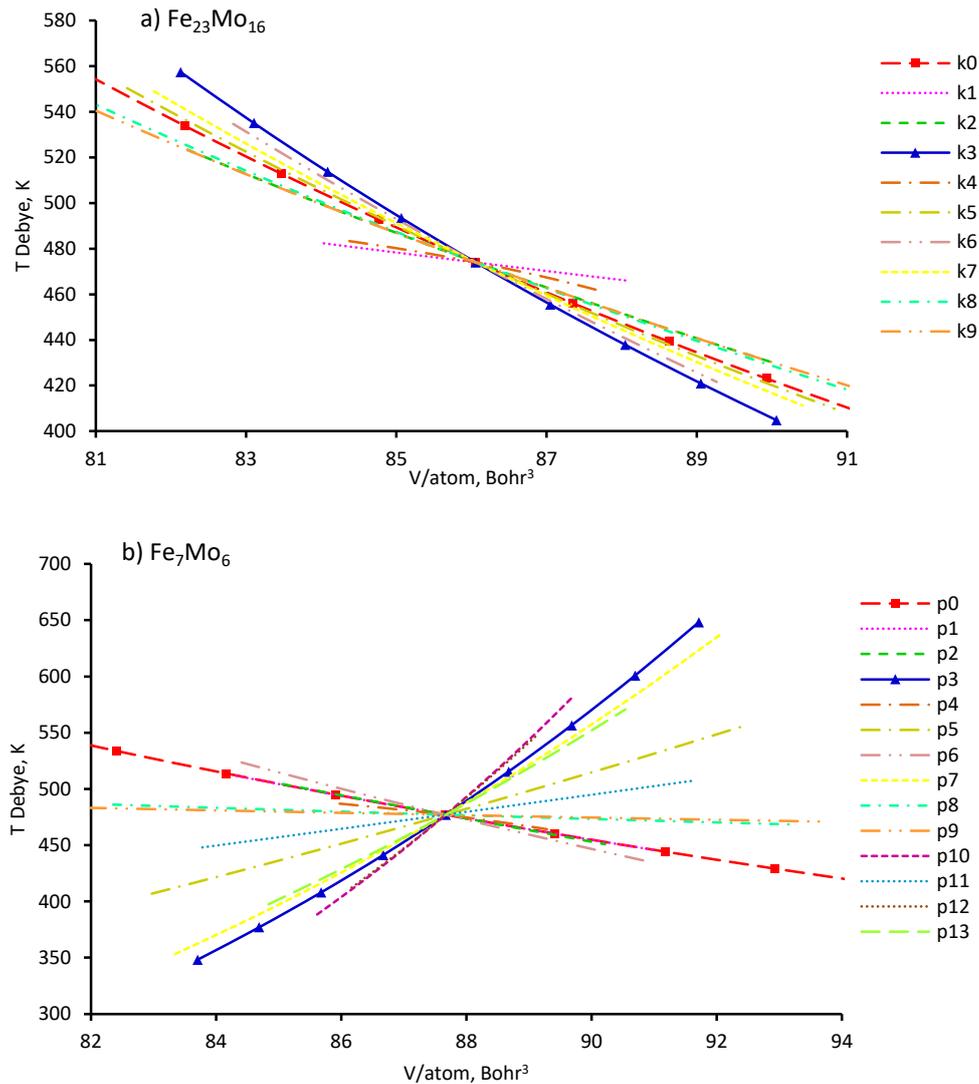

**Figure 5**. Debye temperatures $\theta_D(V)$ as functions of the volume calculated for a) $Fe_{23}Mo_{16}$ along the $k0 \div k9$, and b) $Fe_7Mo_6$ along the $p0 \div p13$ pathways.

To assess the value of the vibrational energy contributed to the free energy of $Fe_{23}Mo_{16}$, the Debye temperatures $\theta_D(V)$ were obtained along the $k0 \div k9$ pathways by Equation (9) and are shown in Figure 5 (a). As shown in this Figure, all the $\theta_D(V)$ functions have negative slopes at the equilibrium volume $V_0$ and at $T = 0K$. The absolute value of the $k3$ slope is the biggest among the other, as shown in Figure 5 (a), therefore, according the QDG approximation the contribution to the free energy along the $k3$ is the largest. Thus, the *k3* has more chances than the other routes to win this virtual competition for the most energetically stable thermal expansion pathway of $Fe_{23}Mo_{16}$.

In order to investigate the influence of vibrational entropy on the stability of $Fe_7Mo_6$ the Debye temperatures $\theta_D(V)$ were calculated using (9) along the *p0 ÷ p13* directions which shown in Figure 5 (b). The $\theta_D(V)$ curves calculated along the *p3, p5, p7, p10, p11, p12* and *p13* pathways have a positive slop, i.e. negative values of the Grüneisen parameter γ, and according the QDG model the expansion along these routes is characterised as having a negative nature. Signs of this phenomenon are reflected in the shape of the total energy curves $E_{stat}(V)$ calculated for these directions, along which the energies increase faster with increasing volume than with decreasing, as shown in Figure 4 (b). The absolute values of the slopes of $\theta_D(V)$ calculated for the *p10* and *p12* routes are higher than those for others, but they have the higher total energies as shown in Figure 4 (b) and this circumstance will diminish the contribution to the free energy calculated along these routes.

The remaining Debye temperature curves, shown in Figure 5 (b), reflect the normal behavior of a solid when heated. The *p0* and *p9* directions have the lowest total energies with increasing volume as shown in Figure 4 (b). Thus, it helps to increase the contribution to the free energy calculated along these routes. Therefore, all these routes will be in competition between themselves for the most energetically favourable thermal expansion pathway of $Fe_7Mo_6$. And it is difficult to predict in advance which direction will have the least energy.

*3.2.3. Accounting for the magnetic subsystems*

It is known that by taking into account the magnetic subsystems can make a significant contribution to the free energy. If the magnetic energy is not taken into account, then the thermal expansion, as was shown in [24, 25] using the Laves $Fe_2Mo$ and μ- $Co_7Mo_6$ phases as an example, will be isotropic or only due to the *c*-parameter, which contradicts the experimental data. Thus, for evaluating the influence of magnetic entropy on the stability of $Fe_{23}Mo_{16}$ and $Fe_7Mo_6$ compounds the local magnetic moments ($\mu_B$) as functions of volumes were obtained for the *k0 ÷ k9* and *p0 ÷ p13* directions and shown in Figure 6 and Figure 7.

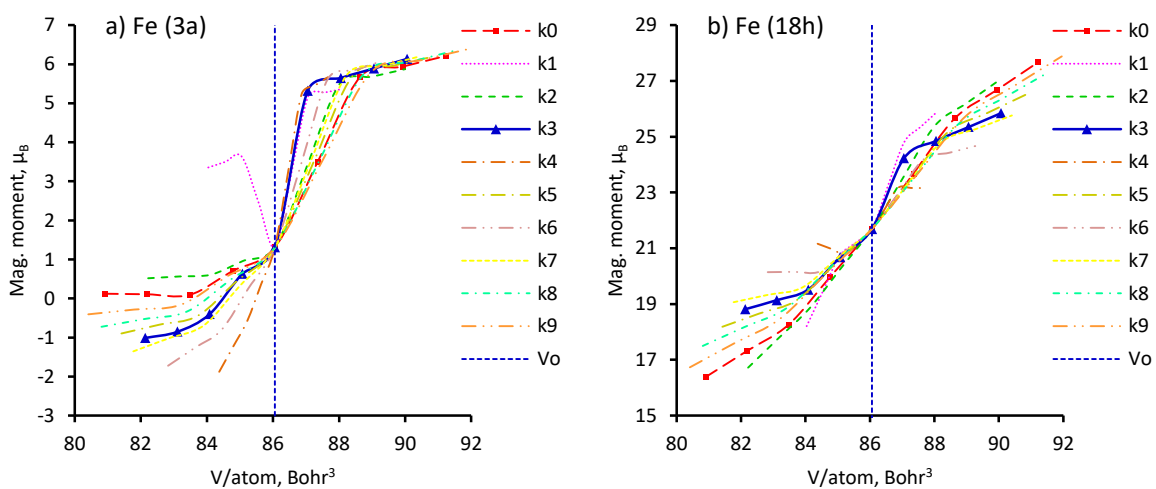

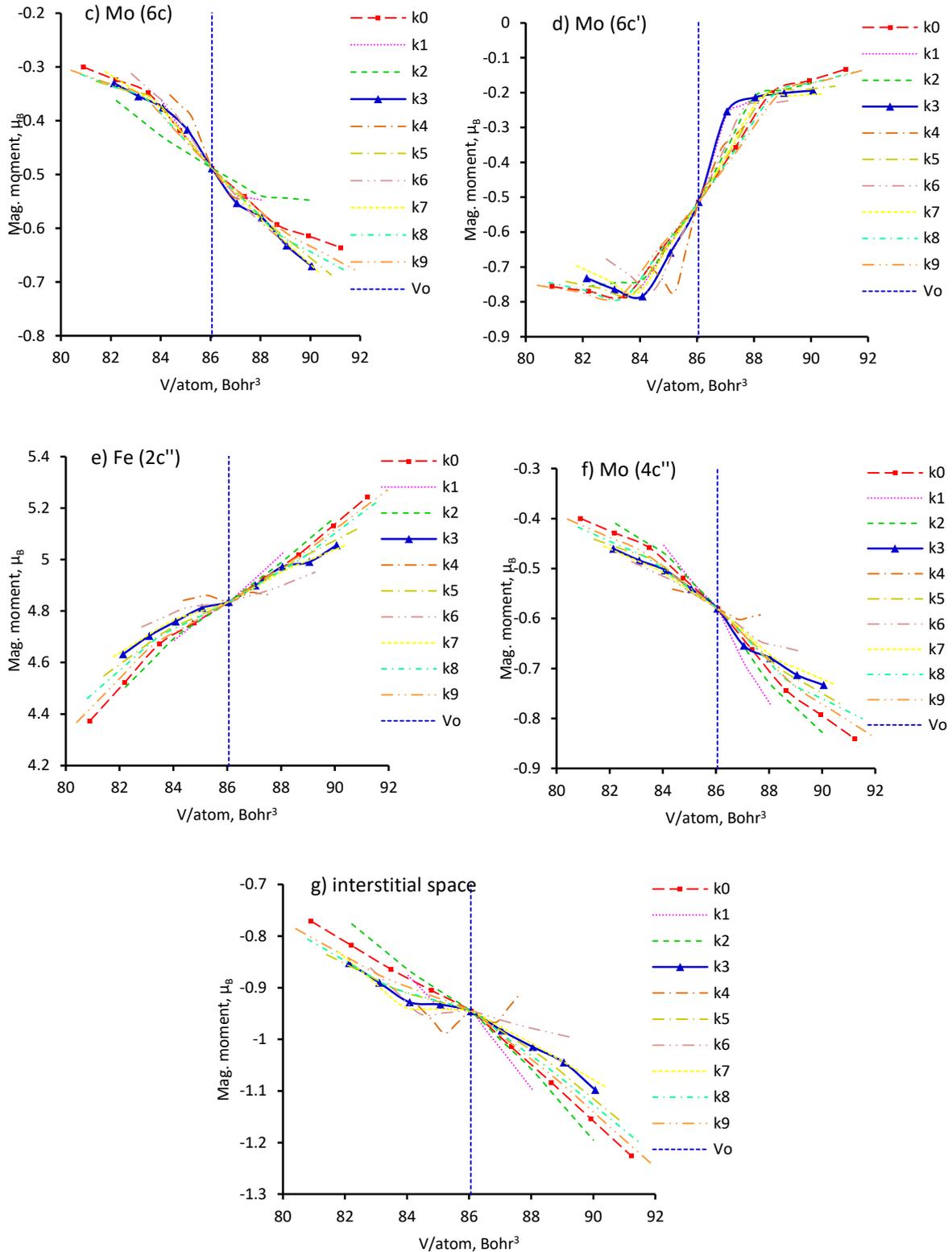

**Figure 6**. The local magnetic moments ($\mu_B$) distributed on the different sub-lattices and in the interstitial space of the $Fe_{23}Mo_{16}$ μ-phase calculated for $k0 \div k9$ pathways: a) Fe atoms on the first sub-lattice, Fe (*3a*); b) Fe atoms on the second sub-lattice, Fe (*18h*); c) Mo atoms on the third sub-lattice, Mo (*6c*); d) Mo atoms on the fourth sub-lattice, Mo (*6c'*); e) two Fe atoms on the fifth sub-lattice, Fe (*2c''*); f) four Mo atoms on the fifth sub-lattice, Mo (*4c''*); g) in the interstitial space. $V_0$ – the equilibrium volume.

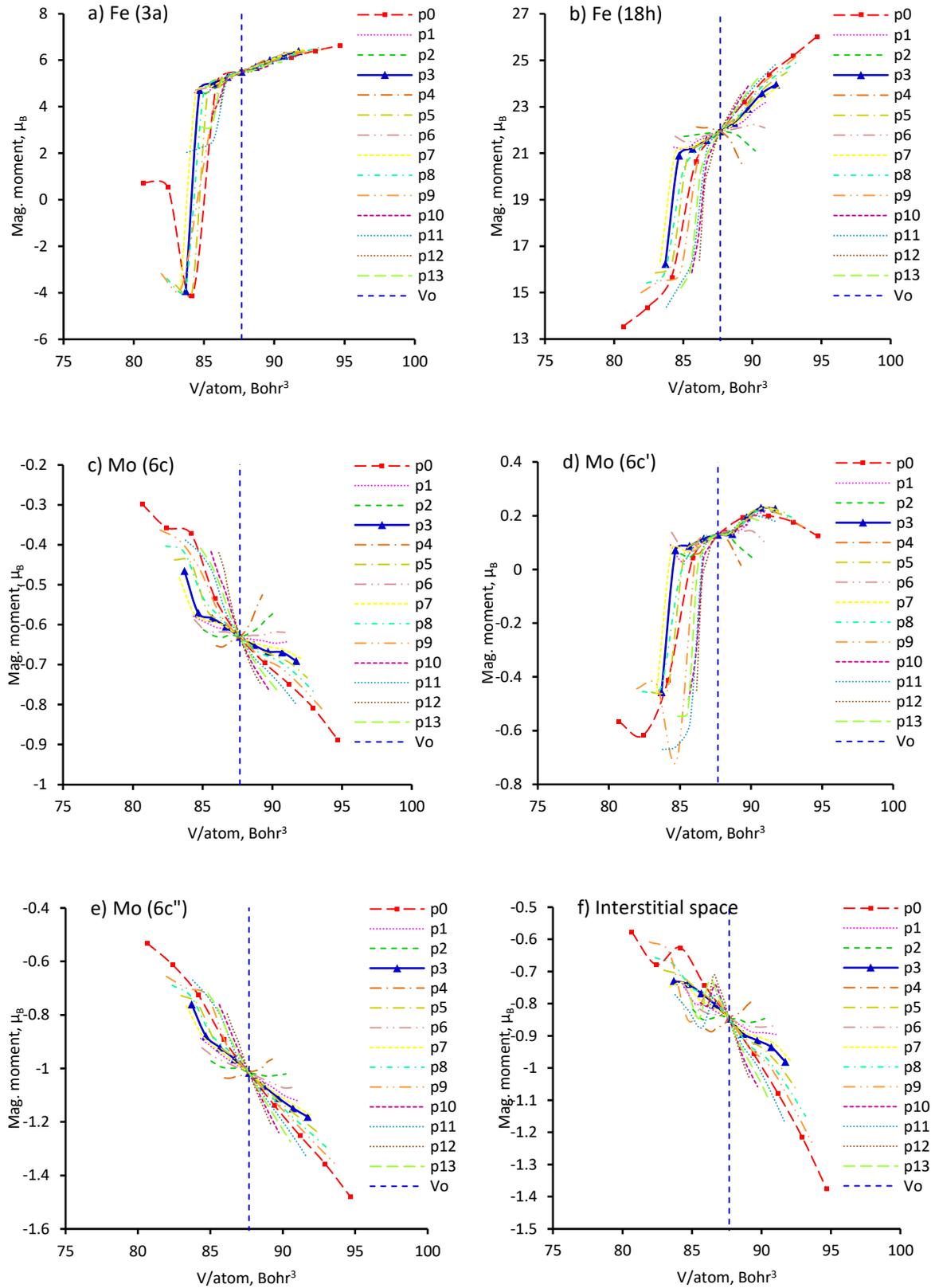

**Figure 7**. The local magnetic moments ($\mu_B$) distributed on the different sub-lattices and in the interstitial space of the Fe$_7$Mo$_6$ μ-phase calculated for *p0 ÷ p13* pathways: a) Fe atoms on the first sub-lattice, Fe (*3a*); b) Fe atoms on the second sub-lattice, Fe (*18h*); c) Mo atoms on the third sub-lattice, Mo (*6c*); d)

Mo atoms on the fourth sub-lattice, Mo (*6c'*); e) Mo atoms on the fifth sub-lattice, Mo (*6c"*); f) in the interstitial space. $V_0$ – the equilibrium volume.

Analyzing the distribution of magnetic moments of atoms over sub-lattices, the following orderings of magnetic moments were obtained for the equilibrium volume $V_0$ associated with T = 0 K. The largest contribution to the magnetic energy of $Fe_{23}Mo_{16}$ is made by the interstitial space, it's about 57 %. This is followed by Fe atoms located on the *18h* sub-lattice, about 30%. The Fe atoms located on the *3a* sub-lattice contribute about 6 %. They are followed by two Fe atoms located on the *2c"* sub-lattice, about 5 %. The rest comes from Mo atoms accommodated on *6c*, *6c'* and *4c"*, these are 0.1%, 1% and 1% respectively.

In the $Fe_7Mo_6$, the alignment is as follows. Fe atoms located on *18h* sub-lattice bring the most to the magnetic energy, it's about 50%. Interstitial space brings 33.5%. The Fe atoms occupied *3a* contribute 11%. This is followed by Mo atoms located on *6c*, *6c'* and *6c"*, they bring 2%, 0.5% and 3% respectively.

In addition, it can be concluded that at T = 0 K the effect of magnetic ordering on the stability of the $Fe_{23}Mo_{16}$ and $Fe_7Mo_6$ is the same.

*3.2.4. Free energy and its components*

The free energies *F(V,T)* of $Fe_{23}Mo_{16}$ and $Fe_7Mo_6$ were calculated by (1) along each *ki* and *pi* directions according to the schemes presented in Figure 3 (a) and Figure 3 (b). The electronic, vibrational and magnetic energy contributions to the *F(V,T)* have been calculated by (2 – 14) at equilibrium volumes $V_0$.

The electronic $F_{el}(V_0(T))$ and vibrational $F_{vib}(V_0(T))$ energies, magnetic entropies $-T \cdot S_{mag}(V_0(T))$ multiplied by -T and the sum of free energies $F(V_0(T))$ calculated by (1 – 14), as temperature functions, along the *k0 ÷ k9* directions for $Fe_{23}Mo_{16}$ are shown in Figure 8, and along the *p0 ÷ p13* for $Fe_7Mo_6$ are shown in Figure 9.

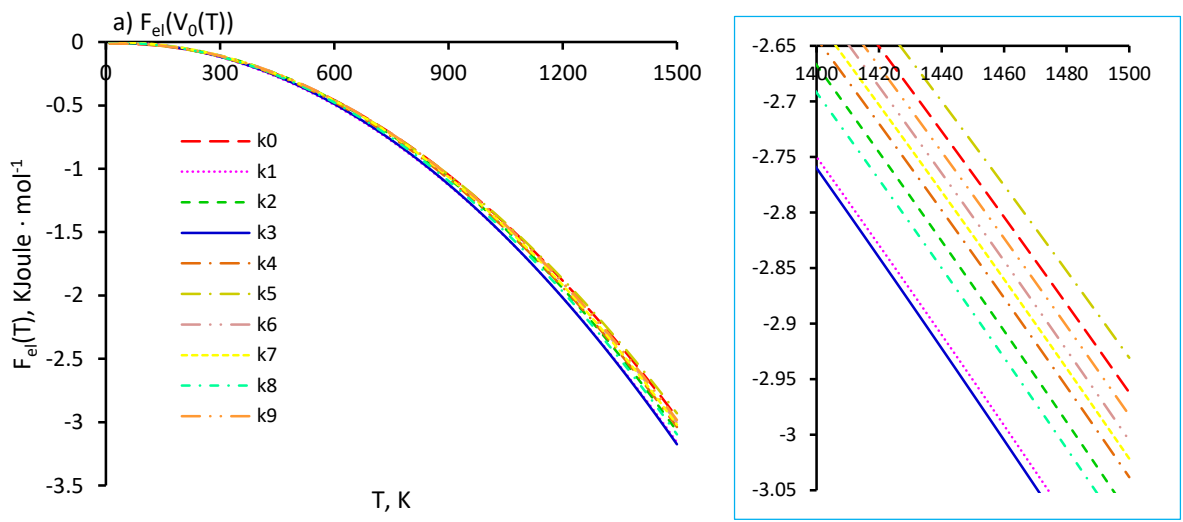

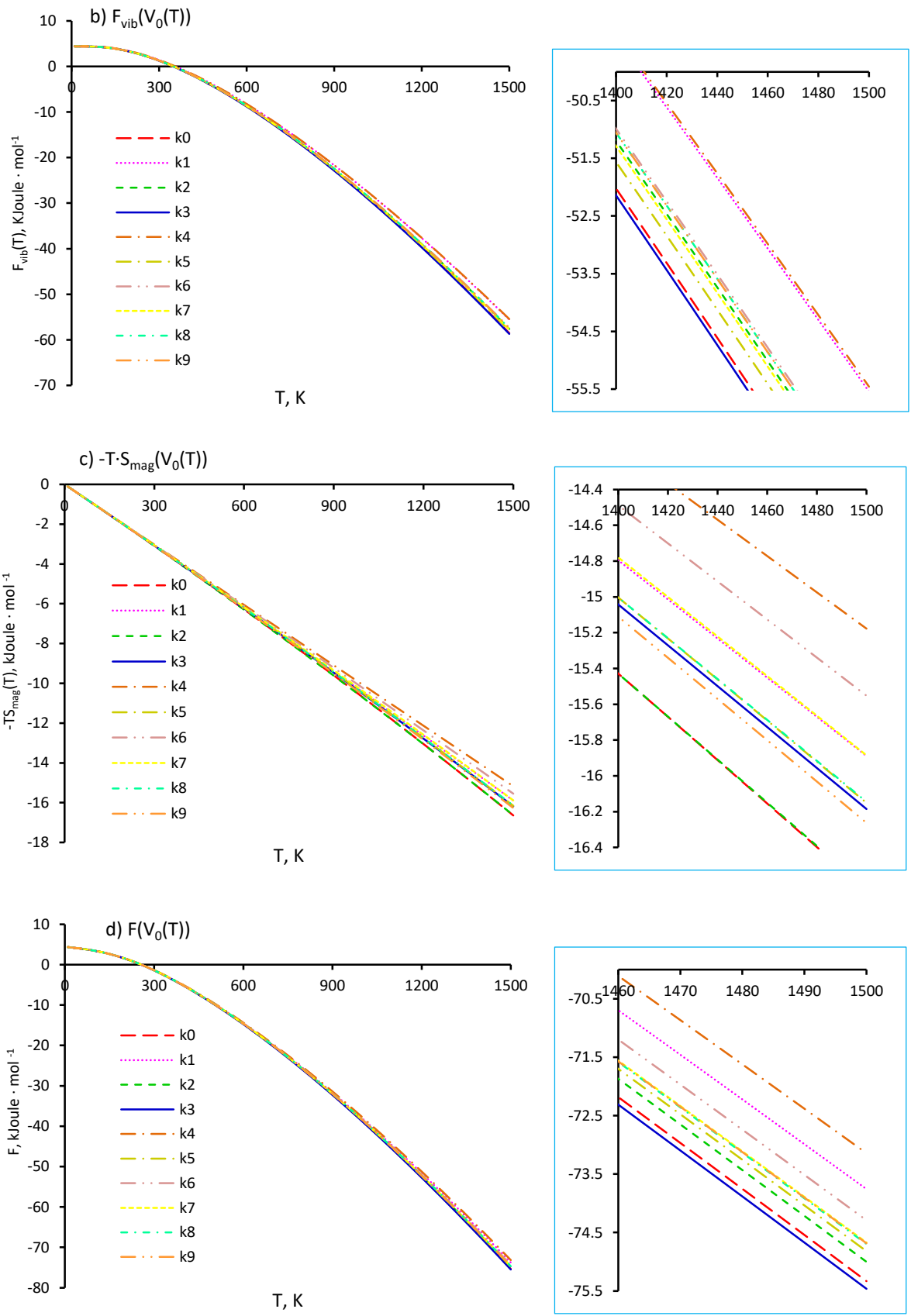

**Figure 8**. Free energy $F(T)$ of $Fe_{23}Mo_{16}$ and its energy constituents calculated for $k0 \div k9$ pathways. a) Electronic energies, $F_{el}(V_0(T))$; b) Vibrational energies, $F_{vib}(V_0(T))$; c) Magnetic entropies, -

$T \cdot S_{mag}(V_0(T))$; d) Free energies, $F(V_0(T))$. On the right sides the magnified values inserted for convenience.

By studying the functions given in Figure 8 we can point out that in the case of $Fe_{23}Mo_{16}$ the electron energy along the *k3* is the lowest electron energy among others as shown in Figure 8 (a). The contribution of vibrational energy to the free energy is larger from the *k3* relative to other directions, as follows from Figure 8 (b). The magnetic contribution to the free energy equally profits from the *k0* and *k2*, as shown in Figure 8 (c).

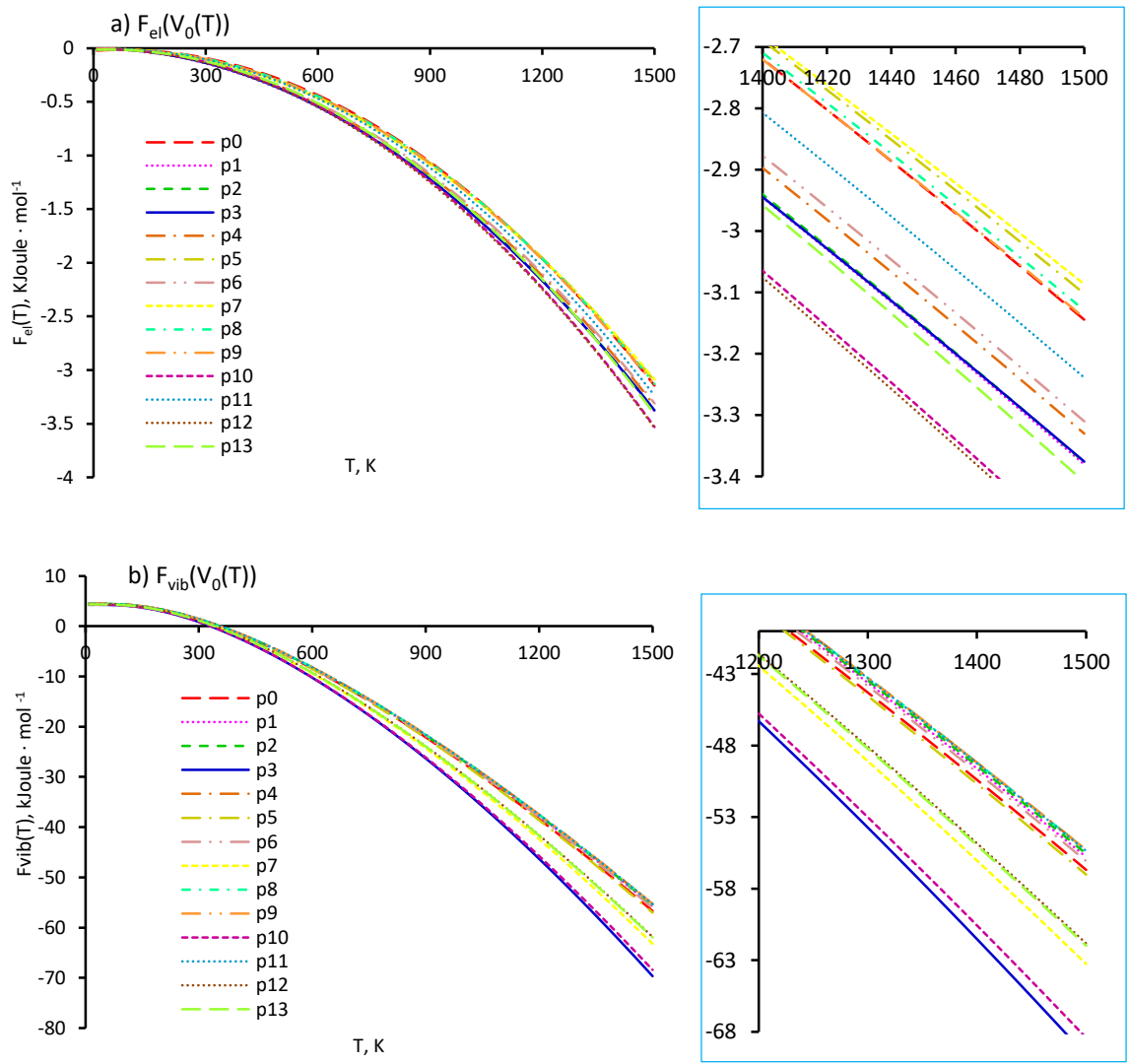

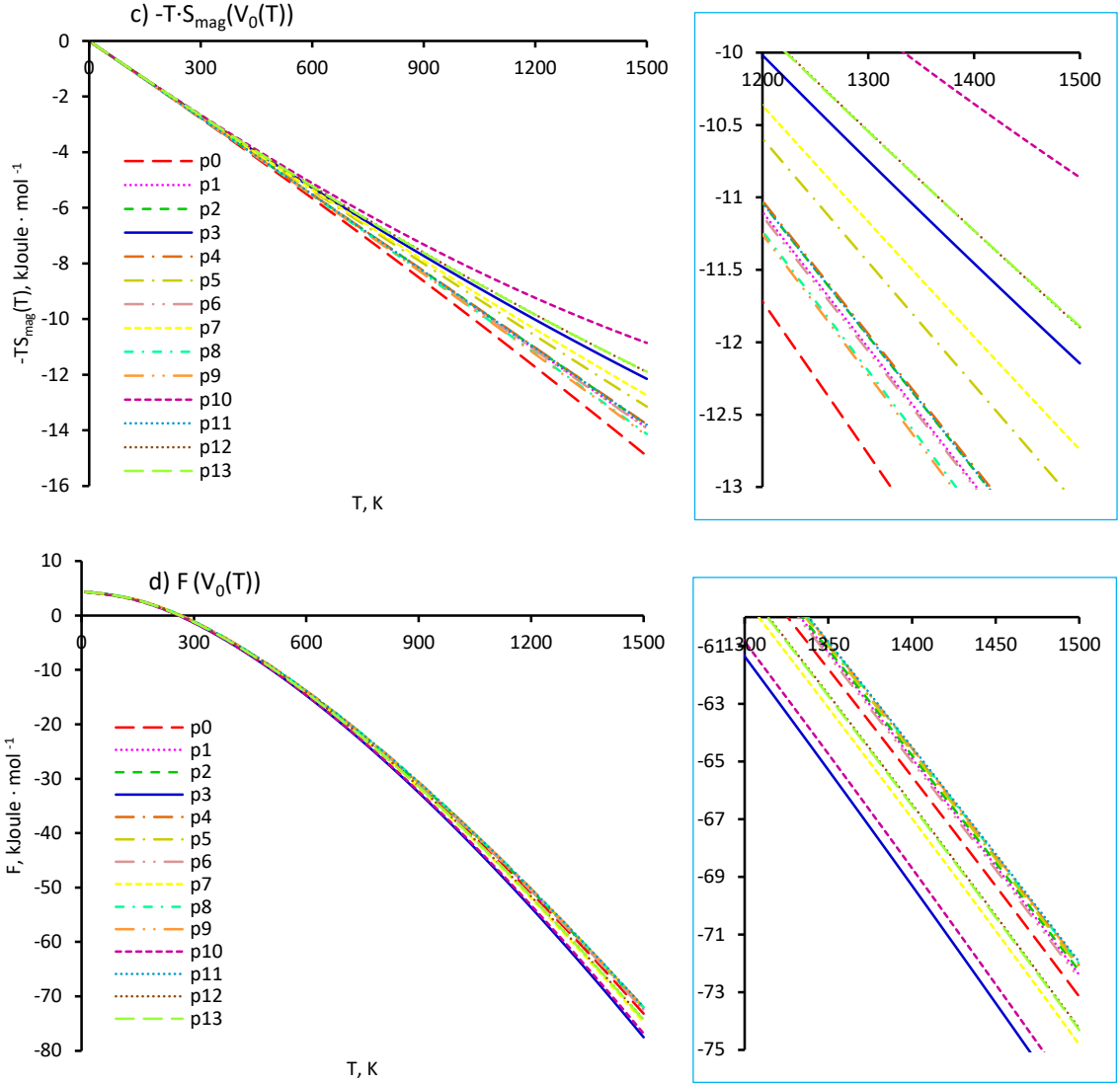

**Figure 9**. Free energy $F(T)$ of $Fe_7Mo_6$ and its energy constituents calculated for $p0 \div p13$ pathways. a) Electronic energies, $F_{el}(V_0(T))$; b) Vibrational energies, $F_{vib}(V_0(T))$; c) Magnetic entropies, -$T \cdot S_{mag}(V_0(T))$; d) Free energies, $F(V_0(T))$. On the right sides the magnified values inserted for convenience.

Considering the graphs related to the $Fe_7Mo_6$, we can note, that the *p12* and *p10* almost equally contribute to the electron energy, as presented in Figure 9 (*a*), more than the other directions. The contribution to the magnetic entropy, as follows from Figure 9 (*c*), is greater from the *p0*. By analyzing the graphs of vibrational subsystem, shown in Figure 9 (*b*), we can conclude that the contribution to the vibrational energy is more significant from the *p3* pathway.

The configurational entropy calculated by Equation (14) in the case of stoichiometry compounds like the $Fe_{23}Mo_{16}$ and $Fe_7Mo_6$, contributes the equal energy for all directions, so it imposes no effect on the comparative analysis.

Therefore, by studying the free energies $F(V_0(T))$ of $Fe_{23}Mo_{16}$ shown in Figure 8 (*d*), we can figure out that the *k3* direction is the most energetically favorable. Therefore, at heating the lattice parameters of $Fe_{23}Mo_{16}$ will be increased from ($a_0$, $c_0$), calculated at T = 0K, towards values on the *k3* shown in Figure 3 (a). Thus, this is the pathway of thermal expansion of $Fe_{23}Mo_{16}$, which is characterized by a negative TEC in the *c* parameter of the lattice.

Comparing the free energies $F(V_0(T))$ of $Fe_7Mo_6$ shown in Figure 9 (*d*), the conclusion can be made, that the *p3* direction is the most energetically stable among the others. It means that at heating the lattice parameters of $Fe_7Mo_6$ will change from ($a_0$, $c_0$), obtained at T = 0K, towards parameters on the *p3* shown in Figure 3 (b). Thus, this is the pathway of negative thermal expansion of $Fe_7Mo_6$.

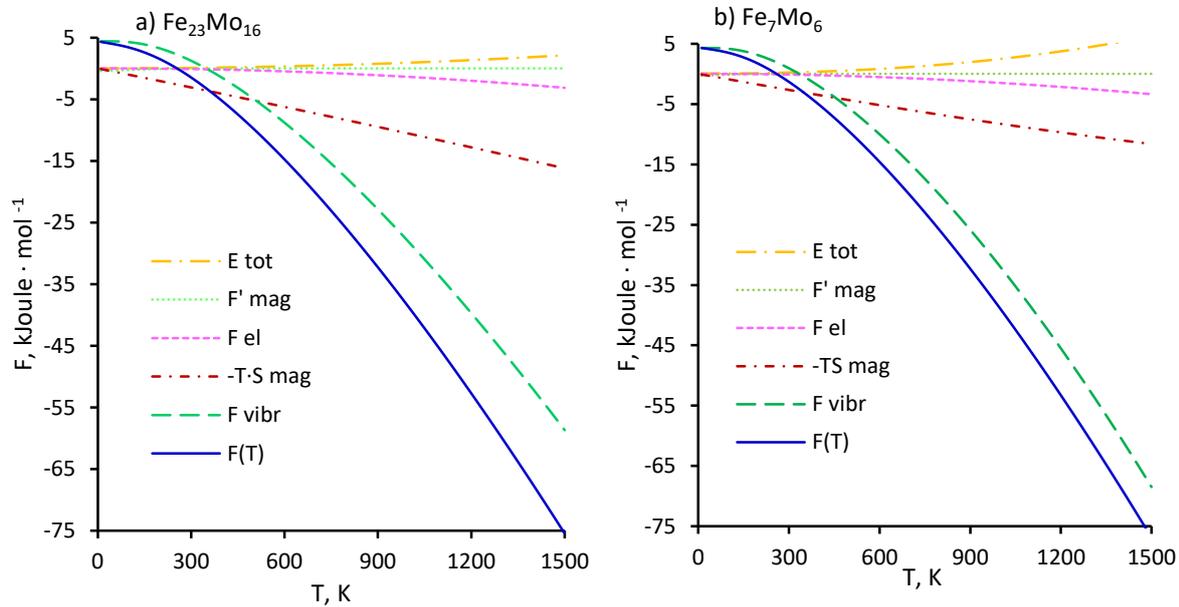

**Figure 10**. The free energies $F(V_0(T))$ for a) $Fe_{23}Mo_{16}$ and b) $Fe_7Mo_6$ with their energy contributions: total, magnetic, electronic, magnetic entropy and vibrational energies calculated along the *k3* and *p3* pathways.

The graphs of free energy $F(V_0(T))$ of $Fe_{23}Mo_{16}$ and $Fe_7Mo_6$ and its energy components: total, magnetic, electronic, vibrational and magnetic entropy, calculated along the *k3* and *p3* pathways are shown in Figure 10 (a) and Figure 10 (b).

Analyzing the calculations, we can find out that at heating the spread of the magnetic energies of $Fe_{23}Mo_{16}$ for different pathways is approximately two times less than that of $Fe_7Mo_6$ as shown in Figure 8 (c) and Figure 9 (c). The effect of electronic energy relative to magnetic on the stability of these compounds is approximately one-fifth for $Fe_{23}Mo_{16}$ and two-thirds for $Fe_7Mo_6$, as follows from Figure 10 (a) and Figure 10 (b). But the main contribution to the free energy of both these compounds is made by vibrational energy. While, all energy contributions should be accurately accounted for correct description of thermodynamic properties.

It is follow from the obtained free energies of $Fe_{23}Mo_{16}$ shown in Figure 8 (d) that the other closest stable direction to the *k3* is the *k0*, and that this compound has a conventional thermal expansion and a negative TEC in parameter *c*.

Thus, in the case of $Fe_{23}Mo_{16}$ the competition takes place between the *k3* and *k0*, where *k0* is the isotropic expansion. This case bears a resemblance with the calculation of $Fe_2Mo$ [24], where the thermal expansion direction competes with the isotropic expansion, too.

Considering the free energies of $Fe_7Mo_6$ calculated along the *p0* ÷ *p13*, shown in Figure 9 (*d*), we can conclude that the other energetically closest pathways to the stable *p3* are the *p10*, *p7*, *p12* and *p13*, which are characterized by negative thermal expansion. Thus, the *p0* and *p9* pathways, which are responsible for the usual thermal behavior of a solid upon heating, lose this peculiar competition.

The free energies $F(V,T)$ of $Fe_{23}Mo_{16}$ and $Fe_7Mo_6$ calculated by (1) at temperature range from 100 K to 1500 K along their thermal expansion pathways $k3$ and $p3$ are shown in Figure 11 (a) and Figure 11(b), respectively. The total static energies calculated by DFT at T = 0 K are shown by red dashed lines, while the equilibrium volumes $V_0(T)$ obtained at each temperature as minimum values of the free energies are shown in these figures by blue dotted lines.

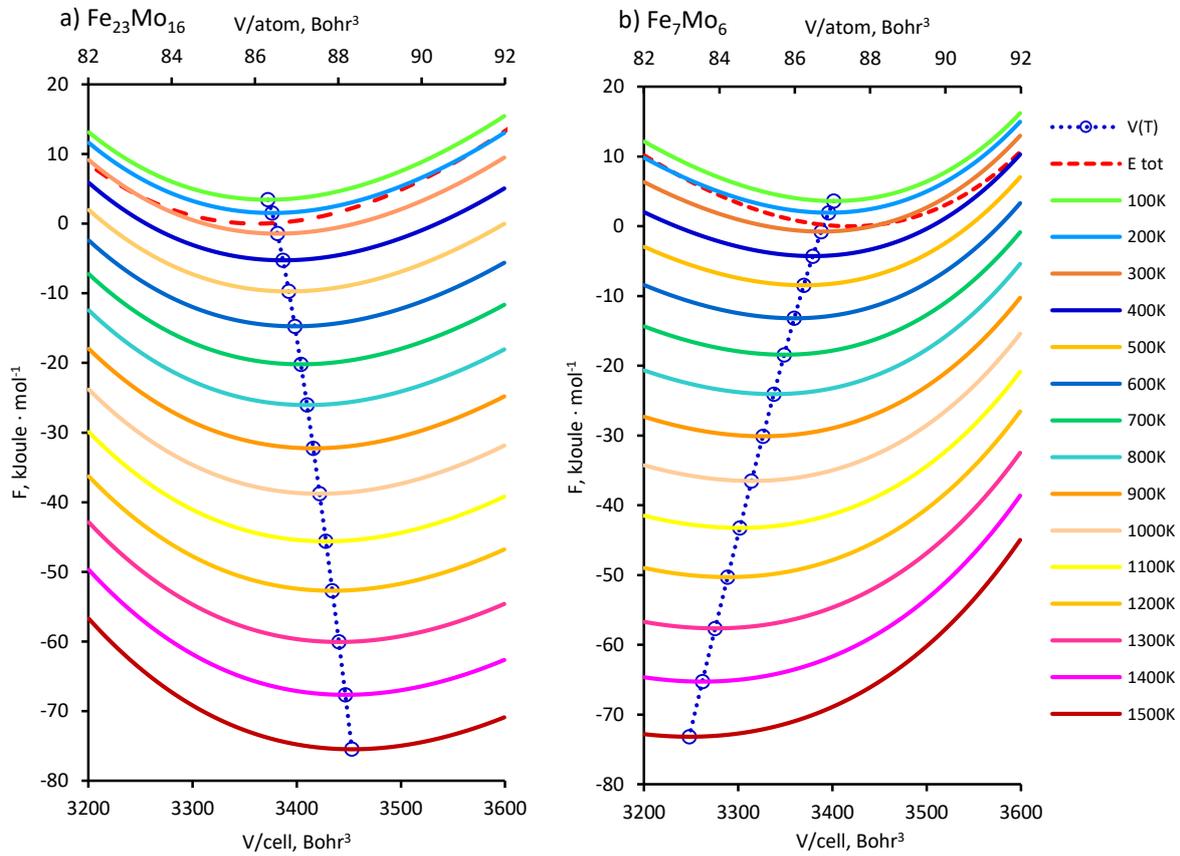

**Figure 11**. Functions of free energy $F(V,T)$ calculated at finite temperatures for a) $Fe_{23}Mo_{16}$ along the $k3$ pathway, and b) $Fe_7Mo_6$ along the $p3$ pathway. The total static energies are shown by the red dashed lines. The equilibrium volumes $V_0(T)$ are presented by the blue dotted lines, open circles indicate the equilibrium volumes.

The thermal expansion of $Fe_{23}Mo_{16}$ can be seen, in Figure 11 (a), as an increase in the equilibrium volume, while the negative thermal expansion of $Fe_7Mo_6$ is a decrease in the volume shown in Figure 11 (b).

## 3.2.5. Thermodynamic properties

Thermal expansion $V(T)$ and heat capacity $Cp(T)$ of $Fe_{23}Mo_{16}$ calculated along the *k3* thermal expansion path are shown in Figure 12 (a) and Figure 12 (b).

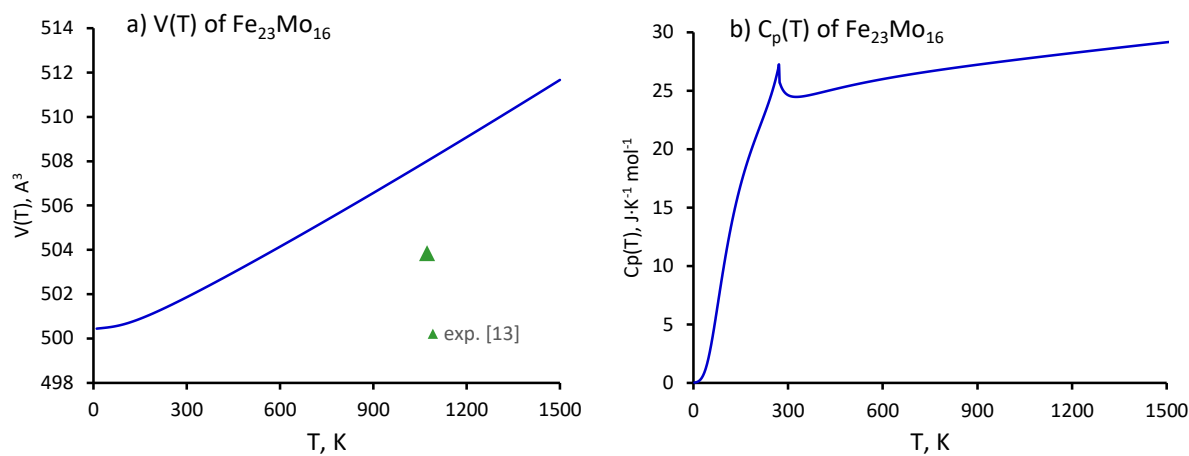

**Figure 12.** a) The volume expansion $V(T)$ of $Fe_{23}Mo_{16}$ calculated along the thermal expansion path is shown together with the experiment data obtained in [13]. b) The heat capacity $Cp(T)$ of $Fe_{23}Mo_{16}$ calculated along the pathway of thermal expansion.

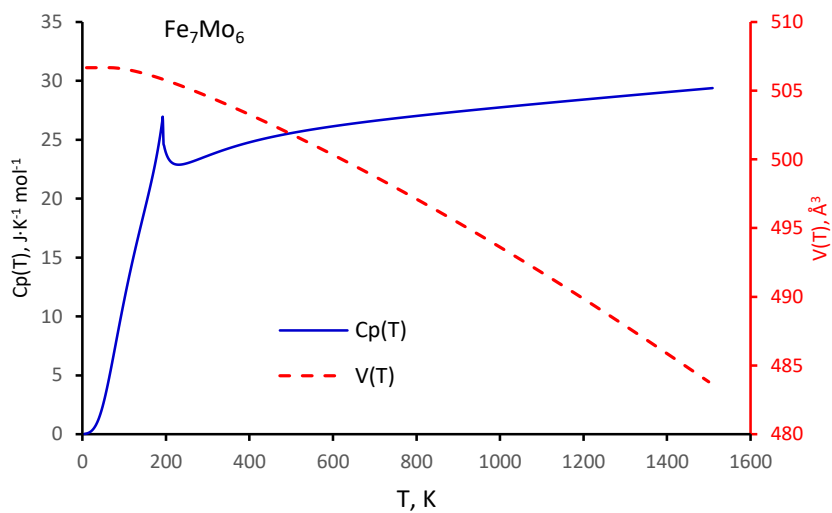

**Figure 13.** The heat capacity $Cp(T)$ of $Fe_7Mo_6$ is shown by the solid blue line (left black axis) calculated along the pathway of negative thermal expansion, while, the volume of negative thermal expansion $V(T)$ of $Fe_7Mo_6$ is shown by the dashed red line (right red axis).

The negative thermal expansion $V(T)$ of $Fe_7Mo_6$, predicted along the *p3* path in this work, is shown in Figure 13 by the red dashed line. The graph of isobaric heat capacity $C_p(T)$ calculated along the negative thermal expansion path *p3* is shown in Figure 13 by the solid blue line. The heat capacity jumps at $T_C$, shown in Figure 12 (b) and Figure 13 are the magnetic phase transition occurring in $Fe_{23}Mo_{16}$ and $Fe_7Mo_6$ upon heating, as the used HJ model [34] predicts. Since the experimental data on the Curie temperature of these compounds are unknown, additional studies are needed. These predicted results may be useful for further research.

*3.3. Discussion*

The quasi harmonic approximation (QHA) nowadays is the most popular approach to obtain thermodynamic properties from first-principles. One of methods used to calculate thermodynamic properties of materials employing the QHA is the Grüneisen formalism. The negative thermal expansion coefficient in-plane of graphite was calculated in the work [41] using QHA and Grüneisen formalism. This formalism involves carrying out phonon calculations, calculating the Grüneisen parameters, and obtaining the TECs from these data. Despite the fact that phonon calculations provide reliable data, their application requires a lot of computer resources.

In this work, the thermal expansions of the stable $Fe_{23}Mo_{16}$ and metastable $Fe_7Mo_6$ μ- phases were predicted by employed QDG approximation and STEP approach. This technique allows to reduce the problem to one-dimensional case and consider the free energy as having only one variable, the volume.

It is known that during heat treatment, the concentration of components in the alloy under study may deviate from the initial concentration as a result of diffusion processes at elevated temperatures, especially if this alloy is metastable. As far as it known, there are no experimental data for the $Fe_{23}Mo_{16}$, but the compositions of $Fe_{23}Mo_{16}$ and $Fe_7Mo_6$ are close to each other. Therefore, it is reasonable to extrapolate the results of calculations for the $Fe_{23}Mo_{16}$ to the experimental data obtained for $Fe_7Mo_6$. Thus, the experimental data [12, 13, 16] obtained for $Fe_7Mo_6$, shown in Figure 3 (a) by orange diamond, green and blue triangles, can be used to assess the lattice parameters of $Fe_{23}Mo_{16}$ calculated at elevated temperatures. These data show that the direction of thermal expansion of $Fe_{23}Mo_{16}$ can indeed be in the vicinity of the *k3* direction, as shown in Figure 3 (a) by the blue line.

The experimental data [13] for $Fe_7Mo_6$ shown in Figure 3 (a) by the red triangle can be considered in the case of $Fe_{23}Mo_{16}$ as some kind of experimental fluctuation, which was not taken into account in this comparative analysis and is presented here for reference.

A comparison of the free energies calculated for different routes showed that the thermal expansion of $Fe_7Mo_6$ μ-phase is negative and not isotropic. The lattice parameters ($a_0$, $c_0$) of $Fe_7Mo_6$ calculated in this work at T = 0 K are close to those calculated theoretically [11], which are shown in Figure 3 (b) by the blue square and listed in Table 1.

The scatter of experimental data [12, 13, 16] shown in Figure 3 (b) seems to reflect the fact that the $Fe_7Mo_6$ compound, as noted above, is a metastable compound at elevated temperatures according to Massalski et al. [6]. So, it is naturally to assume that the diffusion evolving in the compounds is still in a progress, and the lattice parameters are yet changing, and they may end up in the vicinity of the composition of a stable compound such as $Fe_{23}Mo_{16}$. Therefore, in the case of a metastable compound, it is difficult to expect complete agreement with experiment.

However, the experimental data do not contradict the actual theoretical calculations. The experimental parameters obtained in [13] at T = 1073K after annealing for 1500 hours, as shown in Figure 3 (b) by the red triangle, in a case of $Fe_7Mo_6$, can be considered as having a diffusion driving force in the same direction as *p3*. Therefore, a thermal expansion route indeed can lay in the vicinity of the *p3* direction.

It should be noted once again that in the case of metastable compounds it is difficult to rely on experimental data; it is necessary to focus on theoretical calculations. The lattice parameters of the stable $Fe_{23}Mo_{16}$ μ-phase calculated at elevated temperatures are in satisfactorily agreement with the experimental data [12, 13, 16]. In another theoretical work [25], where a thermal expansion of a stable $Co_7Mo_6$ μ-phase was studied using the same approach, good agreement with experimental data is also observed. Thus, it can be assumed that the STEP approach can satisfactorily describe the thermodynamics of the metastable μ-phase. Therefore, the *p3* route can be considered as the predicted pathway of negative thermal expansion of $Fe_7Mo_6$.

It follows from the calculation that the anisotropy of the shape of the total energy surface *E(a, c)*, which allows the *p3* route to have the lowest free energy among other routes, is in a sense responsible for the negative thermal expansion of $Fe_7Mo_6$. And, as consequence, the behavior of the vibrational energy and a negative value of the Grüneisen parameter γ are accountable for the negative coefficients of thermal expansion of $Fe_7Mo_6$. The calculations also show that if we neglect the magnetic entropy, then the path of thermal expansion will remain the same. Therefore, we see that the vibrational energy is the main one in the stabilization of the $Fe_7Mo_6$, while, the magneto-volume effect plays a secondary role. But, in the Laves phase $Fe_2Mo$, magnetic entropy plays a major role [24].

Nevertheless, two additional iron atoms shown in Figure 1 (b) in yellow, which occupy the *6c"* sub-lattice of the $Fe_{23}Mo_{16}$ compound, increase the magnetic entropy contribution according the data presented in Figure 6 (e). They make radical changes in the energy spectrum $F(V_0(T))$, as given in Figure 10 (a), where the role of the magnetic entropy contribution increases, which literally leads to a U-turn of the thermal expansion path by 180 degrees relative to the $Fe_7Mo_6$ compound, as shown in Figure 3 (a) and Figure 3 (b) by *k3* and *p3* blue lines.

So, we can assume that the magnetic subsystem influences the shape of the total energy surface *E(a,c)*, which, in turn, affects the thermal expansion path.

The predicted route *p3* of negative thermal expansion implies that at heating the lattice parameter *c-* of $Fe_7Mo_6$ will increase its value while the parameter *a-* decries and vice versa with decreasing temperature, as shown in Figure 3 (b). However, for the stable $Fe_{23}Mo_{16}$ compound, when heated, the opposite situation occurs along the predicted *k3* pathway, the *a-* parameter increases with increasing temperature, and the *c-* parameter decreases and vice versa, as shown in Figure 3 (a) by blue line. This picture also resembles the thermal expansion trajectory of $Fe_2Mo$ [24, 26], where a negative TEC in parameter *c-* was reported, too. Such thermal behavior can provide stresses in the matrix in both cases, because the thermal expansion of the matrix is usually isotropic. It can reflect the assumption made in [27] that precipitations of μ-phase in superalloys can cause the plastic deformation which contradicts the generally accepted position of brittleness, and it can be considered as confirmation that the hardening (or embrittlement) mechanism caused by the μ-phase is still unknown. Therefore, knowledge of the direction of thermal expansion of the μ-phase can help in understanding the experimental data and the mechanism of material embrittlement. And also in obtaining Gibbs potentials, which can be used in the calculation of phase diagrams.

The purely QDG approximation is considered nowadays as a classical theory and an empirical formalism. But, the resent studies [25, 24] report that the QDG approach together with the STEP method can correctly predict the direction of the thermal expansion of compounds such as the TCP μ- $Co_7Mo_6$ and Laves $Fe_2Mo$ phases, in accordance with experimental data.

Another approximation is phonon calculations of thermodynamic functions which can be carried out for comparison with the calculations performed in this work using the QDG model. These calculations should theoretically coincide with each other.

Such a similar comparison is reported in [26], where the direction of thermal expansion of $Fe_2Mo$ Laves phase obtained based on phonon spectra was compared with the direction calculated using the QDG approach, and it shows a good agreement between these two methods.

The analysis presented in [26] shows that the vibrational energy calculated from the phonon spectra agrees well with the energy calculated as the sum of the Debye energy and the magnetic entropy.

The $Fe_2Mo$ and μ- phases have the similar TCP hexagonal structures, thus the outcome of this analysis can be extrapolated to the more complex μ-phases.

It is known, the phonon calculations already include magnetic contributions, so it is difficult to estimate them separately. This is an advantage of this method, i.e. we can separately assess the influence of electronic, vibrational and magnetic energies on a phase stability. Thus, it can give us a complete picture of the stabilizing factors that affect compounds.

The main results of studies [24 - 26] using the STEP method are as follows. i) The shape of the energy surface *ΔE(a,c)* of a compound reflects posable directions of thermal expansions. ii) The Debye temperatures calculated along different paths may give us a glance which trajectory having the more chances of being the most energetically favorable path of thermal expansion. iii) Nevertheless, only the knowledge of anisotropies of the surface energy shape together with the electronic and magnetic entropies can provide a correct prediction of thermodynamic functions.

These results are consistent with the results of the present work. Namely, the Debye temperature curves calculated for different directions of thermal expansion really reflect the shape of the energy surfaces of $Fe_{23}Mo_{16}$ and $Fe_7Mo_6$, as follows from Figure 4 and Figure 5. And in order to correctly calculate the thermodynamics of compounds, along with vibrational energy it is necessary to carefully take into account the electronic and magnetic entropies. Therefore, it can be concluded that this method can successfully describe the thermodynamic properties of TCP Laves and µ- phases.

It necessary to note, that QDG approach is consuming less computer time than methods based on the calculations of phonon spectra. As complexity of crystal structures grows, this difference in the computer time required for calculations increases. Therefore, calculating the thermodynamic properties of compounds using the QDG approximation and STEP method can be advantageous.

Finally, this study predicts that the thermal expansion direction of the intermetallic $Fe_{23}Mo_{16}$ µ-phase of the Fe-Mo system exhibits the same non-isotropic behavior as the $Fe_2Mo$ Laves phase reported in [24]. Therefore, it can be assumed that these compounds have similar physical factors affecting their stability.

## 4. Conclusions

The finite-temperature DFT calculations have been carried out to calculate electronic, vibrational and magnetic free energy contributions for the stable $Fe_{23}Mo_{16}$ and metastable $Fe_7Mo_6$ µ-phases. By employing the approach of searching a thermal expansion path of compounds the thermal expansion pathway of $Fe_{23}Mo_{16}$, with a negative TEC in parameter *c*, is correctly calculated in accordance with the available experimental data. The negative thermal expansion is predicted in $Fe_7Mo_6$ which is a metastable compound. It is found that the surface energy anisotropy and the negative value of Grüneisen parameter are responsible for this effect. The nature of thermal expansion of both these compounds is not isotropic. The available experimental data of lattice parameters which are close in the composition to the $Fe_{23}Mo_{16}$ µ-phase don't show contradiction with the results of this work. Thus, the approach of searching a thermal expansion path of compounds may be used to study thermodynamic properties of the metastable µ-phase. Two additional iron atoms, which occupy the *6c"* sub-lattice of $Fe_{23}Mo_{16}$ compound make dramatically change in the direction of thermal expansion relatively the $Fe_7Mo_6$. The elastic constants, bulk modulus, elastic sound velocities, Debye and Curie temperatures were calculated at ground state. The work shows that the vibrational energy is the main factor influencing on the stability of $Fe_{23}Mo_{16}$ and $Fe_7Mo_6$ compounds, and that all energy contributions should be equally accounted for obtaining correct thermodynamic properties of µ-phase.

**Acknowledgments**

The research was financially supported by the Russian Foundation for Basic Research as a part of scientific project № 19-03-00530.

# Appendix

**Table A1.** The set of distortion matrices $D_i$ together with the equations reflecting changes in total energies $\Delta E(\delta)$ of hexagonal lattice, resulted after applied strain, used to calculate the elastic constants $C_{ij}$ of deformation tensors for Fe$_{23}$Mo$_{16}$ and Fe$_7$Mo$_6$ µ-phases.

| $D_i$ | $\Delta E(\delta)$ |
|---|---|
| $D_1 = \begin{pmatrix} 1+\delta & 0 & 0 \\ 0 & 1 & 0 \\ 0 & 0 & 1 \end{pmatrix}$ | $\Delta E = V_0(\tau_1\delta + \frac{C_{11}}{2}\delta^2)$ |
| $D_2 = \begin{pmatrix} \frac{1+\delta}{(1-\delta^2)^{1/3}} & 0 & 0 \\ 0 & \frac{1-\delta}{(1-\delta^2)^{1/3}} & 0 \\ 0 & 0 & \frac{1}{(1-\delta^2)^{1/3}} \end{pmatrix}$ | $\Delta E = V_0\left[(\tau_1-\tau_2)\delta + \frac{1}{2}(C_{11}+C_{22}-2C_{12})\delta^2\right]$ |
| $D_3 = \begin{pmatrix} \frac{1+\delta}{(1-\delta^2)^{1/3}} & 0 & 0 \\ 0 & \frac{1}{(1-\delta^2)^{1/3}} & 0 \\ 0 & 0 & \frac{1-\delta}{(1-\delta^2)^{1/3}} \end{pmatrix}$ | $\Delta E = V_0\left[(\tau_1-\tau_3)\delta + \frac{1}{2}(C_{11}+C_{33}-2C_{13})\delta^2\right]$ |
| $D_4 = \begin{pmatrix} 1 & 0 & 0 \\ 0 & 1 & 0 \\ 0 & 0 & 1+\delta \end{pmatrix}$ | $\Delta E = V_0(\tau_3\delta + \frac{C_{33}}{2}\delta^2)$ |
| $D_5 = \begin{pmatrix} \frac{1}{(1-\delta^2)^{1/3}} & 0 & 0 \\ 0 & \frac{1}{(1-\delta^2)^{1/3}} & \frac{\delta}{(1-\delta^2)^{1/3}} \\ 0 & \frac{\delta}{(1-\delta^2)^{1/3}} & \frac{1}{(1-\delta^2)^{1/3}} \end{pmatrix}$ | $\Delta E = V_0(2\tau_4\delta + 2C_{44}\delta^2)$ |

**Table A2.** Ground state elastic compliances $S_{ij}$ (TPa$^{-1}$) of single crystals of Fe$_7$Mo$_6$ and Fe$_{23}$Mo$_{16}$.

| Compound | $S_{11}$ | $S_{12}$ | $S_{13}$ | $S_{33}$ | $S_{44}$ |
|---|---|---|---|---|---|
| Fe$_7$Mo$_6$ | 3.276 | -1.019 | -1.434 | 6.484 | 9.657 |
| Fe$_{23}$Mo$_{16}$ | 3.304 | -1.017 | -1.421 | 6.441 | 10.119 |

A hexagonal crystal of µ-phase defined by Bravais lattice in a matrix form was expressed as

$$\boldsymbol{R} = \begin{pmatrix} \frac{\sqrt{3}}{2} & \frac{1}{2} & 0 \\ -\frac{\sqrt{3}}{2} & \frac{1}{2} & 0 \\ 0 & 0 & 1 \end{pmatrix} \quad (A1)$$

The polycrystalline elastic properties of a compound, such as bulk modulus $B$, shear modulus $G$, Young's modulus $E$, and Poisson's ratio $v$, were estimated by Voigt-Reuss-Hill (VRH) approximation according to [37]

$$B = \frac{1}{2}(B_V + B_R), G = \frac{1}{2}(G_V + G_R) \quad (A2)$$

$$E = \frac{9GB}{3B+G}, v = \frac{3B-E}{6B} = \frac{E-2G}{2G} \quad (A3)$$

Where the bulk and shear modules calculated from the Voigt model are denoted as $B_V$ and $G_V$, while $B_R$ and $G_R$ identify the Reuss model.

The moduli of the μ-phase can be obtained from the symmetry of the hexagonal lattice and calculated as follows

$$B_V = \frac{1}{9}(2(C_{11} + C_{12}) + 4C_{13} + C_{33}) \quad (A4)$$

$$B_R = \frac{(C_{11}+C_{12})C_{33}-2C_{13}^2}{C_{11}+C_{12}+2C_{33}-4C_{13}} \quad (A5)$$

$$G_V = \frac{1}{30}(C_{11} + C_{12} + 2C_{33} - 4C_{13} + 12C_{44} + 12C_{66}) \quad (A6)$$

$$G_R = 15/(14S_{11} + 4S_{33} - 8S_{13} - 10S_{12} + 6S_{44}) \quad (A7)$$

Where $C_{ij}$ and $S_{ij}$ are elastic constants and compliances components of the tensor and its inverse matrix, respectively.

The Debye temperature $\theta_D$ was calculated according to [42]

$$\theta_D = \frac{h}{k_B}\left[\frac{3n}{4\pi}\left(\frac{N_A\rho}{M}\right)\right]^{1/3} V_m \quad (A8)$$

An averaged sound velocity $V_m$ in polycrystalline aggregate was formulated as

$$V_m = \left[\frac{1}{3}\left(\frac{2}{v_t^3} + \frac{1}{v_l^3}\right)\right]^{-1/3} \quad (A9)$$

The $v_t$ transverse and $v_l$ longitudinal elastic wave velocities were calculated in accordance with

$$v_t = \left(\frac{G}{\rho}\right)^{1/2}, \quad v_l = \left(\frac{(B+4G/3)}{\rho}\right)^{1/2} \quad (A10)$$

where $\rho$ is a density of the compound.

The velocities of elastic waves propagating along [001] direction were calculated as follows [43]

$$V_L = \left(\frac{C_{33}}{\rho}\right)^{1/2}, \quad V_{T1} = V_{T2} = \left(\frac{C_{44}}{\rho}\right)^{1/2} \quad (A11)$$

While, the velocities of elastic waves propagating along [100] direction were given as

$$V_L = \left(\frac{C_{11}}{\rho}\right)^{1/2}, \quad V_{T1} = \left(\frac{C_{66}}{\rho}\right)^{1/2}, \quad V_{T2} = \left(\frac{C_{44}}{\rho}\right)^{1/2} \quad (A12)$$

where $V_L$ and $V_T$ are longitudinal and transverse elastic wave velocities.